\documentclass[12pt]{article}
\usepackage{latexsym,epsfig,graphicx,amsmath,amssymb,amscd,multirow,chicago,psfrag,float,pdfpages,array,booktabs,appendix,paralist, subfigure, titlesec}
\usepackage{subfig}
\newcommand{\thetavec}{{\boldsymbol{\theta}}}

\newcommand{\Sigmavec}{{\boldsymbol{\Sigma}}}

\newcommand{\betavec}{{\boldsymbol{\beta}}}

\newcommand{\gammavec}{{\boldsymbol{\gamma}}}

\newcommand{\E}{\mathbf{E}}
\newcommand{\D}{\mathcal{D}}

\newcommand{\Var}{{\rm Var}}
\newcommand{\Corr}{{\rm Corr}}

\newcommand{\sigmahat}{\widehat{\sigma}}
\newcommand{\betahat}{\widehat{\beta}}
\newcommand{\gammahat}{\widehat{\gamma}}
\newcommand{\rhohat}{\widehat{\rho}}
\newcommand{\thetavechat}{\widehat{\thetavec}}
\newcommand{\gammavechat}{\widehat{\boldsymbol{\gamma}}}

\newcommand{\wh}{\widehat}

\newcommand{\xvec}{\boldsymbol{x}}
\newcommand{\yvec}{\boldsymbol{y}}
\newcommand{\Xmat}{\textbf{X}}

\newcommand{\degreeC}{{^\circ\mathrm{C}}}
\newcommand{\Sigmavechat}{\widehat{\Sigmavec}}
\newcommand{\Rvec}{\boldsymbol{R}}
\newcommand{\argmax}{\operatornamewithlimits{argmax}}

\begin{document}
\title{Semi-parametric Models for Accelerated Destructive Degradation Test Data Analysis}

\author{Yimeng Xie$^1$,  Caleb B. King$^1$, Yili Hong$^1$, and Qingyu Yang$^2$\\[2ex]
{\small $^1$Department of Statistics, Virginia Tech, Blacksburg, VA 24061}\\
{\small $^2$Department of Industrial and Systems Engineering, Wayne State University, Detroit, MI 48202}
}

\date{}

\maketitle

\begin{abstract}
Accelerated destructive degradation tests (ADDT) are widely used in industry to evaluate materials' long term properties. Even though there has been tremendous statistical research in nonparametric methods, the current industrial practice is still to use application-specific parametric models to describe ADDT data. The challenge of using a nonparametric approach comes from the need to retain the physical meaning of degradation mechanisms and also perform extrapolation for predictions at the use condition. Motivated by this challenge, we propose a semi-parametric model to describe ADDT data. We use monotonic B-splines to model the degradation path, which not only provides flexible models with few assumptions, but also retains the physical meaning of degradation mechanisms (e.g., the degradation path is monotonically decreasing). Parametric models, such as the Arrhenius model, are used for modeling the relationship between the degradation and accelerating variable, allowing for extrapolation to the use conditions. We develop an efficient procedure to estimate model parameters. We also use simulation to validate the developed procedures and demonstrate the robustness of the semi-parametric model under model misspecification. Finally, the proposed method is illustrated by multiple industrial applications.


\end{abstract}
\newpage

\section{Introduction}
\subsection{Motivation} \label{sec: Background}

It is important for manufacturers to understand the lifetime of their products in order to ensure accurate marketing and determine areas for improvement. While lifetime testing is the most common approach, for many materials it is more informative to observe the degradation of some performance characteristic, such as the tensile strength of an adhesive bond, over time. The lifetime is determined by a ``soft failure" when the characteristic drops below a predetermined level. This form of testing is known as degradation testing.

Several varieties of degradation testing have been developed to accommodate unique circumstances. Due to the long service life of many new materials, degradation testing under normal use conditions is often not feasible. By exposing the material to a more harsh environment, such as higher levels of temperature or humidity compared to the use conditions, degradation data can be collected more efficiently. Thus, an accelerating variable is often used in degradation tests. In some applications, measurements of the degradation level are destructive. That is, the units being tested are destroyed or the physical characteristics changed in a significant manner. An example could be determining the strength of a material by measuring the force needed to break it. This form of testing, combined with an accelerating variable, is referred to as accelerated destructive degradation testing (ADDT). Because of the nature of the testing, ADDT must be analyzed differently from other common forms of degradation testing, such as repeated-measures degradation testing (RMDT), in which multiple measurements can be taken from the same unit.

Current procedures for analyzing ADDT data involve an assumed parametric model for the degradation path over time and a parametric form for the accelerating-variable effect. The predominance of parametric models is mainly due to the need for extrapolation in two aspects; extrapolation in time and extrapolation to the use conditions. For example, an ADDT may cover only 100-70\% of the original material's strength and be performed at an elevated temperature range ($60-80\degreeC$), but interest lies at strengths 50-70\% of the original at a temperature of $30\degreeC$. These parametric models tend to be material-specific and at present there seems to be no general model that can be applied to a wide variety of materials.

Even though there has been tremendous statistical research in nonparametric methods, the current industrial practice is still to use application-specific parametric models to describe ADDT data.  Motivated by multiple industrial applications, we aim to bridge this gap between the statistical research and current industrial practice. Instead of a case-by-case parametric modeling approach, we propose a general and flexible semi-parametric model to describe ADDT data. The challenge of using a nonparametric approach comes from the need to retain the physical meaning of degradation mechanisms and performing extrapolations for predictions at the use condition. To overcome those challenges, the semi-parametric model consists of a nonparametric model for the degradation path and a parametric form for the accelerating-variable effect. In order to preserve the monotonic nature of many degradation paths, the nonparametric model portion will be constructed based on monotonic spline methods. For the parametric model portion, commonly used models, such as the Arrhenius relationship for temperature, will be used for extrapolation. Parameter estimation and inference procedures will also be developed.

\subsection{Related Literature} \label{sec: Related Literature}
The literature on accelerated degradation data modeling and analysis can be divided into two areas: RMDT and ADDT. In the pioneering work, \citeN{LuMeeker1993} used RMDT data to estimate failure-time distribution via the framework on mixed-effects models. \citeN{MeekerEscobarLu1998} introduced nonlinear mixed-effects models for RMDT data, which were derived from physical-failure mechanisms. Introductory level description of degradation models can be found in \shortciteN{Gorjianetal2010}, and \citeN{MeekerHongEscobar2011}. \citeN{YeXie2015} provided a comprehensive review of the state-of-art methods in modeling RMDT data.

In the area of ADDT data modeling and analysis, \citeN[Chapter~11]{Nelson1990} used ADDT data from an insulation to estimate performance degradation. \shortciteN{Escobaretal2003} provided a parametric model and method to analyze the ADDT data collected from an adhesive bond. \shortciteN{Tsaietal2013} considered the problem of designing an ADDT with a nonlinear model motivated by a polymer dataset. \citeN{LiDoganaksoy2014} used a parametric model to model ADDT data collected from a temperature accelerated test to study the degradation of seal strength. In all existing methods for analyzing ADDT data, the parametric method is the most popular.

Compared to parametric models of degradation data, spline functions tend to be more flexible and require less assumptions regrading the model formulation. Because the degradation path is often monotonic in nature, monotone splines are suitable for modeling degradation paths. \citeN{Ramsay1988} suggested using a basis of I-splines (integrated splines) for semi-parametric modeling. \citeN{HeShi1998} considered the use of B-splines with $L_{1}$ optimization. \citeN{Meyer2008} extended the work in \citeN{Ramsay1988} by proposing cubic monotone splines. \citeN{LeitenstorferTutz2007} considered the use of monotone B-splines in generalized additive models. For other applications of monotone B-splines, one can refer to \citeN{KanungoGayHaralick1995} and \citeN{Fengler2014}. In addition, \citeN{EilersMarx1996} proposed a flexible class of P-splines. \citeN{BollaertsEilersMechelen2006}, \citeN{Hofneretal2011}, and \citeN{HofnerKneibHothorn2014} considered the estimation of monotonic effects with P-splines.

Related to RMDT models, \shortciteN{Yeetal2014} considered semi-parametric estimation of Gamma processes. \shortciteN{Hongetal2015}, and \citeN{XuHongJin2015} used shape-restricted splines to model the effects of time-varying covariates on the degradation process. There is little literature, however, on the use of semi-parametric models in ADDT data modeling and analysis.






\subsection{Overview} \label{sec: Overview}
The rest of this paper is organized as follows. Section~\ref{sec: The Data and Model} introduces some general notation for ADDT data. It also presents in detail the construction of the semi-parametric model using monotonic B-splines. In Section \ref{sec: The Estimation Procedure}, we present a procedure for estimating the unknown parameters as well as procedures for conducting inference on ADDT data based on this model. We conduct simulation studies in Section \ref{sec: Simulation Study} to investigate the performance of the semi-parametric method with special consideration of model misspecification. In Section \ref{sec: Applications}, we apply the model to data from several published datasets and provide comparisons with other well-known parametric models. Finally, Section \ref{sec: Conclusion and Areas for Future Work} contains conclusions and areas for future research.

\section{The Semi-parametric Model} \label{sec: The Data and Model}
\subsection{General Setting}\label{sec: data}
Let $y_{ijk}$ be the degradation measurement for the $k$th sample at level $i$ of the accelerating variable $\mathcal{AF}_i$ and the $j$th observation time point $t_{ij}$, $i=1,\ldots, I$, $j=1,\ldots,J_i$, and $k = 1, \ldots, n_{ij}$, where $n_{ij}$ denotes the sample size at $t_{ij}$. Let $n=\sum_{i=1}^{I}\sum_{j=1}^{J_i}n_{ij}$ be the total number of observations. A general form of the degradation model is
\begin{align} \label{eqn: degradation model}
y_{ijk}=\D(t_{ij}, x_i; \thetavec)+\varepsilon_{ijk},
\end{align}
where $x_i=h(\mathcal{AF}_{i})$ is a function of the accelerating variable, $\thetavec$ is a vector of unknown parameters in the degradation path, and $\varepsilon_{ijk}$ is an error term that describes unit-to-unit variability. For the purposes of illustration, we will assume that the degradation path is monotone decreasing with time. The model can easily be generalized to paths that are increasing with time. We will also be considering temperature as the accelerating factor as it is the most common form of acceleration encountered in ADDT. However, the model can easily incorporate other types of acceleration, such as voltage.


For temperature-accelerated processes, the Arrhenius model is often used to describe the relationship between degradation and temperature. This model uses a transformed temperature level given as
\begin{align}\label{eqn: Arrhenius.model}
x_i=\frac{-11605}{\textrm{Temp}_{i} + 273.16}.
\end{align}
Here, $\textrm{Temp}_{i}$ is in degrees Celsius, and 11605 is the reciprocal of the Boltzmann's constant (in units of eV). The value 273.16 in the denominator is used to convert to the Kelvin temperature scale.

\subsection{The Scale Acceleration Model} \label{sec: model}
We propose the following semi-parametric functional forms for the degradation model in \eqref{eqn: degradation model}.
\begin{align}\label{eqn: semi-parametric degradation model}
&\D(t_{ij}, x_i; \thetavec)= g\left[\eta_{i}(t_{ij};\beta);\gammavec\right],  \\\label{eqn:scale.acc}
&\eta_{i}(t; \beta) =\frac{t}{\exp{(\beta s_i)}},\quad s_i =x_{\max}-x_i,\\\label{eqn:error.dist}
&\varepsilon_{ijk} \sim \textrm{N}(0, \sigma^2),\quad\textrm{and}\quad \Corr(\varepsilon_{ijk}, \varepsilon_{ijk'})=\rho,\,\, k \ne k'.
\end{align}
Here, $g(\cdot)$ is a monotone decreasing function with unknown parameter vector $\gammavec$,  $\beta$ is an unknown parameter associated with the accelerating variable, and $\thetavec=(\gammavec', \beta, \sigma, \rho)'$ is the vector containing all of the unknown parameters. The quantity $x_{\max}=-11605/[\max_{i}{(\textrm{Temp}_{i})} + 273.16]$ is defined to be the transformed value of the highest level of the accelerating variable.

The model in \eqref{eqn: semi-parametric degradation model} falls within the class of scale acceleration models. For a specific stress level $i$, $\D(t, x_i; \thetavec)$ is a decreasing function of time $t$, in which $\beta$ controls the degradation rate through time-scale factor $\exp{(\beta s_i)}$ in \eqref{eqn:scale.acc}. A smaller time-scale factor corresponds to a rapid decrease in degradation. When the acceleration level is at its highest, $s_{\max}=x_{\max}-x_{\max}=0$. In this case, $\eta_{i}(t; \beta)=t$ implies that the degradation path no longer relies on $\beta$, and $$\D(t, x_{\max}; \thetavec)= g(t;\gammavec).$$
Thus, the function $g(\cdot)$ can be interpreted as the baseline degradation path for the scale acceleration model in \eqref{eqn: semi-parametric degradation model}. The distribution of error terms $\varepsilon_{ijk}$ are specified in \eqref{eqn:error.dist} with parameters $\sigma$ and $\rho$. In particular, we consider a compound symmetric correlation structure for measurements taken on the same temperature and time point. Measurements at different temperatures and times are assumed to be independent.

Let $y_{M}$ be the lowest degradation level present in the observed data. Then the scale-acceleration model and the monotonicity of $g(\cdot)$ will allow one to extrapolate the degradation level to $y_{M}$ for any given acceleration level.  Let $\D_f$ be the failure threshold. Then, if $y_M<D_f$, one can use the semi-parametric model to obtain failure information at the use conditions through this extrapolation. This is particularly useful since, in general, measurements may be available below $\D_f$ for only some of the highest levels of the accelerating variable. In fact, some industrial standards require that tests be run until the degradation level drops below $\D_f$ for several acceleration levels. However, extrapolation beyond $y_{M}$ is not possible due to the nonparametric construction of the $g(\cdot)$, which is the tradeoff for this kind of model flexibility.

\subsection{Nonparametric Form for Baseline Degradation Path} \label{sec: Monotonic B-splines}
We use nonparametric methods to estimate the baseline degradation path $g(\cdot)$. Specifically, we use monotonic B-splines to model the baseline degradation path. This not only provides flexible models, but also retains the physical meaning of degradation mechanisms (e.g., the degradation path is monotonically decreasing).

Consider a set of interior knots $d_1 \le \cdots \le d_{N}$, and two boundary points $d_0$ and $d_{N+1}$. The entire set of ordered knots are
$$
d_{-q}=\cdots=d_{0} \le d_1 \le \cdots \le d_{N} \le d_{N+1} = \cdots = d_{N+q+1},
$$
where the lower and upper boundary points are appended $q$ times and $q$ is the polynomial degree.  For notational simplicity, we rewrite the subscripts in the ordered knot sequences as $d_{1}, \cdots, d_{N+2q+2}$. The total number of basis functions is $p=N+q+1$. The $l$th B-spline basis function of degree $q$ evaluated at $z$ can be recursively obtained in the following formulas:
\begin{gather*}
B_{0,  l}(z) = \mathbf{1}(d_{l} \le z < d_{l+1}), \\
B_{q,  l}(z) = \frac{z-d_{l}}{d_{l+q}-d_{l}} B_{q-1, l}(z) + \frac{d_{l+q+1}-z}{d_{l+q+1}-d_{l+1}} B_{q-1,  l+1}(z),
\end{gather*}
where $l=1,\cdots, p$, and $\mathbf{1}(\cdot) \textrm{ is an indicator function}$. The degradation model can then be expressed as
\begin{gather} \label{eqn: semi-parametric degradation model, B-splines}
y_{ijk}=\sum_{l=1}^{p} \gamma_{l} B_{q, l}[\eta_{i}(t_{ij};\beta)]+\varepsilon_{ijk},
\end{gather}
where $\gamma_{l}$'s are the coefficients.



To ensure the degradation path is monotone decreasing, we require the first derivative of $\D(\tau_{ij}, x_i; \thetavec)$ be negative. For B-spline basis functions, \citeN{Boor2001} proved that the derivative of $\D(t, x_i; \thetavec)$ with respect to $\eta_i(t;\beta)$ is
$$
\frac{d\D(t, x_i; \thetavec)}{d\eta_i(t;\beta)}= \sum_{l=2}^{p} (q-1) \frac{(\gamma_{l} - \gamma_{l-1}) }{d_{l+q+1}-d_{l}} B_{q-1, l}[\eta_{i}(t;\beta)].
$$
As B-spline basis functions are nonnegative, it follows that $\gamma_{l} \leqslant \gamma_{l-1}$ for all $2 \le l \le p$ gives a sufficient condition for a monotone decreasing degradation path. However, except for basis functions with degree $q=1, 2$, it is not a necessary condition. \citeN{FritschCarlson1980} derived the necessary conditions for cubic splines ($q=3$), though for higher order splines necessary conditions are as yet unclear.

\section{Estimation and Inference}\label{sec: The Estimation Procedure}
\subsection{Parameter Estimation}
Let $\yvec_{ij}=(y_{ij1},\ldots, y_{ij n_{ij}})'$, $\boldsymbol{\varepsilon}_{ij}=(\varepsilon_{ij1},\ldots, \varepsilon_{ijn_{ij}})'$, $\yvec  = (y_{11}',\ldots, y_{1J_{1}}',  \ldots, y_{I1}',\ldots, y_{IJ_{I}}')'$, $\boldsymbol{\varepsilon}  = (\varepsilon_{11}',\ldots, \varepsilon_{1J_{1}}',  \ldots, \varepsilon_{I1}',\ldots, \varepsilon_{IJ_{I}}')'$ and $\gammavec = (\gamma_1, \ldots, \gamma_p)'$. The degradation model in (\ref{eqn: semi-parametric degradation model, B-splines}) can be written as
\begin{equation} \label{eqn: degradation model in matrix model}
\yvec  = \Xmat_{\beta}\gammavec + \boldsymbol{\varepsilon},
\end{equation}
where
\begin{equation*}
\Xmat_{\beta} = \begin{bmatrix}
 B_{q, 1}[\eta_{1}(t_{11};\beta)] & \cdots & B_{q, p}[\eta_{1}(t_{11};\beta)]\\
 B_{q, 1}[\eta_{1}(t_{12};\beta)] & \cdots & B_{q, p}[\eta_{1}(t_{12};\beta)]\\
 \vdots & \ddots & \vdots \\
 B_{q, 1}[\eta_{I}(t_{IJ_I};\beta)] & \cdots & B_{q, p}[\eta_{I}(t_{IJ_I};\beta)]
\end{bmatrix} ,
\end{equation*}
and $\boldsymbol{\varepsilon} \sim \mathrm{N}\left(\boldsymbol{0}, \Sigmavec \right)$. Here, $\Sigmavec= \textrm{Diag} \left(\Sigmavec_{11},   \ldots, \Sigmavec_{1 J_{1}},  \ldots, \Sigmavec_{I1},   \ldots, \Sigmavec_{I J_{I}} \right)$ and $\Sigmavec_{ij}=\sigma^2 [ (1-\rho) I_{n_{ij}} $$ + \rho J_{n_{ij}} ]$, where $I_{n_{ij}}$ is an $n_{ij} \times n_{ij}$ identity matrix and $J_{n_{ij}}$ is an $n_{ij} \times n_{ij}$ matrix of 1's. We can also rewrite $\Sigmavec=\sigma^2 \Rvec$, where $\Rvec=\textrm{Diag} \left(\Rvec_{11}, \ldots, \Rvec_{1 J_1}, \ldots, \Rvec_{I1},\ldots,\Rvec_{I J_I} \right)$ and $R_{ij}=(1-\rho) I_{n_{ij}} $$ + \rho J_{n_{ij}}$.

We use likelihood-based methods to estimate the unknown parameters $\thetavec=(\gammavec', \beta, \sigma, \rho)'$. For now, we consider estimation of $\thetavec$ with a given number of knots and knot locations. We will give a discussion on knot selection in Section~\ref{sec: Knots Selection}. A particular challenge to the estimation comes from the constraints on $\gammavec$, namely that $\gamma_l \le \gamma_{l-1}, 2 \le l \le p$. We also note that, for a given $\beta$, $\Xmat_{\beta}$ is known, in which case \eqref{eqn: degradation model in matrix model} becomes a linear model with a correlated covariance structure. Thus, we proceed by first deriving estimates of $(\gammavec', \sigma, \rho)'$ given $\beta$ and then use a profile likelihood approach to estimate $\beta$.

The estimates of $\gammavec$ and $(\sigma, \rho)'$ are obtained using an iterative procedure. In particular, at the $m$th iteration, given estimates $(\sigmahat^{(m-1)}, \rhohat^{(m-1)})'$, the value of $\gammavechat^{(m)}$ is obtained by minimizing
\begin{gather}
Q(\gammavec)=(\yvec-\Xmat_{\beta}\gammavec)'  \left(\Sigmavechat ^{(m-1)} \right)^{-1} (\yvec-\Xmat_{\beta}\gammavec) \nonumber \\
\textrm{subject to } \gamma_{l} \leqslant \gamma_{l-1}, 2 \le l \le p.  \label{eqn: quadratic object function for gamma}
\end{gather}
Equation (\ref{eqn: quadratic object function for gamma}) is a quadratic object function with linear constraints and so can be solved with quadratic programming techniques. Given $\gammavechat^{(m)}$, one can then obtain $(\sigmahat^{(m)}, \rhohat^{(m)})'$ using restricted maximum likelihood (REML) so long as $\gammavechat^{(m)}$ does not take values on the boundary of the linear constraints. If the solution of equation (\ref{eqn: quadratic object function for gamma}) does take values on the boundary of the linear constraints, we can still consider approximate REML to obtain these estimates. Let $\gammavechat_{u}^{(m)}$ represent all of the unique values in $\gammavechat^{(m)}$ and $p_{u}$ be the length of $\gammavechat_{u}^{(m)}$. For each unique value $\gammahat_{i, u}^{(m)}$, let $\xvec_{i, \beta u}$ be the sum of the corresponding columns in $\Xmat_{\beta}$. Then we have $\Xmat_{\beta}\gammavechat^{(m)}=\Xmat_{\beta u}\gammavechat_{u}^{(m)}$, where $\Xmat_{\beta u}=(\xvec_{1, \beta u}, \cdots, \xvec_{p_{u}, \beta u})$. The approximate REML log-likelihood is then
\begin{gather}
\mathcal{L}_{\textrm{REML}}(\sigma, \rho|\gammavechat^{(m)})=-\frac{1}{2} \left\{ \log |\Sigmavec| +\log|\Xmat_{\beta u}' \Sigmavec^{-1} \Xmat_{\beta u} | +(\yvec-\Xmat_{\beta}\gammavechat^{(m)})' \Sigmavec^{-1}(\yvec-\Xmat_{\beta}\gammavechat^{(m)})  \right\}.
\label{eqn: REML loglikelihood for covariance parameter}
\end{gather}
The covariance parameter estimates $(\sigmahat^{(m)}, \rhohat^{(m)})'$ are those values that maximize equation (\ref{eqn: REML loglikelihood for covariance parameter}). In particular, after some calculation it can be shown that $\sigmahat^{^{(m)}}$ has the following closed-form expression
$$
\sigmahat^{^{(m)}}= \left[\frac{(\yvec-\Xmat_{\beta}\gammavechat^{(m)})' (\widehat{\Rvec}^{(m-1)})^{-1}(\yvec-\Xmat_{\beta}\gammavechat^{(m)})}{n-p_{u}}\right]^{\frac{1}{2}}.
$$
Thus, $\rhohat^{(m)}$ can be obtained from a one dimensional optimization problem. That is,
\begin{align*}
\rhohat^{(m)} = \argmax_{\rho}&\left\{ - \log |(\sigmahat^{^{(m)}})^{2}\Rvec| - \log|(\sigmahat^{^{(m)}}) ^{-2}\Xmat_{\beta u}' \Rvec^{-1} \Xmat_{\beta u} | \right. \\
&  \left.-(\sigmahat^{^{(m)}})  ^{-2}(\yvec-\Xmat_{\beta}\gammavechat^{(m)})' \Rvec^{-1}(\yvec-\Xmat_{\beta}\gammavechat^{(m)})\right\} .
\end{align*}
Upon convergence, the estimates of $(\gammavechat', \sigmahat, \rhohat)'$ are obtained for a given $\beta$, denoted by $(\gammavechat_{\beta}', \sigmahat_{\beta}, \rhohat_{\beta})'$. The initial values $(\sigmahat^{(0)}, \rhohat^{(0)})'$ can be easily obtained by fitting a non-constrained model.

The profile log-likelihood for $\beta$ is given as
\begin{equation*}
\mathcal{L}(\beta, \gammavechat_{\beta}, \sigmahat_{\beta}, \rhohat_{\beta}) = \log{\left\{\frac{1}{\sqrt{2\pi}|\widehat{\Sigmavec}_{\beta}|^{1/2}}\exp{\left[-\frac{(\yvec -\Xmat_{\beta}\gammavechat_{\beta})\widehat{\Sigmavec}_{\beta}^{-1}(\yvec -\Xmat_{\beta}\gammavechat_{\beta})}{2}\right]}\right\}}.
\end{equation*}
In practice, one can first estimate $(\gammavec', \sigma, \rho)'$ for a specified range of values of $\beta$, then compute $\mathcal{L}(\beta, \gammavechat_{\beta}, \sigmahat_{\beta}, \rhohat_{\beta})$ as a function of $\beta$. The estimate $\betahat$ is the value that maximizes this function. The final estimates are denoted by $\thetavechat=(\gammavechat', \betahat, \sigmahat, \rhohat)'$.


Once the model parameters have been estimated, other parameters related to reliability can then be estimated. For example, the mean time to failure (MTTF), denoted by $m_f$, is one of many ways to evaluate the reliability of a product/material. Based on the semi-parametric model, we can derive an estimate $\wh{m}_f$ at a use condition $x_f$ and failure threshold $\D_f$ by solving
\begin{align*}
\sum_{l=1}^{p} \gammahat_{l} B_{q, l}\left(\frac{\wh{m}_f}{\exp[\wh\beta(x_{\max}-x_{f})]}\right)=\D_f.
\end{align*}

\subsection{Spline Knots Selection} \label{sec: Knots Selection}
The number of knots and knot locations are a key component to using B-splines to model the degradation path. In addition, it is also necessary to determine the maximum degree of the B-splines. For knot selection, we first fix the degree of the B-splines and then find the optimum knot locations. Optimality is determined by a variation of the Akaike information criterion:
\begin{gather} \label{eqn: AIC}
\textrm{AIC}=-2 \log{\left\{\frac{1}{\sqrt{2\pi}|\widehat{\Sigmavec}|^{1/2}}\exp{\left[-\frac{(\yvec -\Xmat_{\betahat}\gammavechat)\widehat{\Sigmavec}^{-1}(\yvec -\Xmat_{\betahat}\gammavechat)}{2}\right]}\right\}}+2\times edf,
\end{gather}
where $edf$ is the effective degrees of freedom in $\gammavec$ plus three for the parameters $(\beta, \sigma, \rho)'$. \citeN{WangMeyerOpsomer2013} and \citeN{Meyer2012} discussed constrained spline regression for both independent and correlated error cases. In particular, they showed how to calculate the effective degrees of freedom for a constrained fit through the use of a cone projection, which is the trace of the projection matrix. Because we have $p-1$ linear constraints, the effective degrees of freedom in $\gammavec$ has a value from $1$ to $p$, where $p$ corresponds to a unconstrained fit. Letting $q$ denote the degree of the B-spline functions, the procedure for knot selection is as follows:

\begin{enumerate}
\item Determine the optimum number of interior knots $N_{\textrm{opt}, q}$ which minimizes the AIC. The default knot locations are equally-spaced sample quantiles. That is, if number of interior knots is $N$, the default knot locations are $b/N, b=1, \cdots, N-1$.
\item Delete each of the internal knots in sequence. The knot whose deletion leads to the greatest reduction in AIC is removed. Repeat until no more existing knots can be removed.
\end{enumerate}

The whole procedure is to be repeated for different B-spline degrees until the optimal knot sequence is determined. This knot selection procedure is similar to the procedure in \citeN{HeShi1998}. The sample size for an ADDT is typically small and so a low degree of spline ($q \leqslant 4$) and a small number of interior knots ($1 \leqslant  N \leqslant 5$) are usually sufficient to provide a good fit to the data.

\subsection{Statistical Inference} \label{sec: Statistical Inference}
Inference based on the semi-parametric model in (\ref{eqn: degradation model in matrix model}) can rely on either asymptotic theory or a bootstrap procedure.  Because the bootstrap method is straightforward and easy to implement, we use a nonparametric bootstrap to calculate confidence intervals (CI) for the parameters and pointwise CI for the degradation path. The error term in model (\ref{eqn: degradation model in matrix model}) can be written as
$$\varepsilon_{ijk}=u_{ij}+e_{ijk},$$
where $u_{ij} \sim \textrm{N}(0, {\sigma}_{u} ^ {2}), e_{ijk} \sim \textrm{N}(0, {\sigma}_{e} ^ {2})$, $\Corr(u_{ij}, e_{ijk})=0$, ${\sigma}_{u} ^ {2}=\rho\sigma^{2}$, and ${\sigma}_{e} ^ {2}=(1-\rho)\sigma^{2}$.  That is, the error term in model (\ref{eqn: degradation model in matrix model}) can be written as the sum of a random effect term $u_{ij}$ and an independent error term $e_{ijk}$. To obtain the CI, one could resample from the estimated random effect term $\widehat{u}_{ij}$ and the estimated independent error term $\widehat{e}_{ijk}$ separately. However, \citeN{CarppenterGoldsteinrasbash2003} showed that directly resampling from $\widehat{u}_{ij}$ and $\widehat{e}_{ijk}$ will cause bias. Therefore, we adjust $\widehat{u}_{ij}$ and $\widehat{e}_{ijk}$ prior to bootstrapping. That is,
$$\widehat{u}_{ij}^{c}=\left[\sum_{ij}\widehat{u}_{ij}^{2}/(nJ_{n})\right]^{-1/2}\widehat{\sigma}_{u}\widehat{u}_{ij},\quad \textrm{and} \quad \widehat{e}_{ijk}^{c}=\left[\sum_{k}\widehat{e}_{ijk}^{2}/(n_{ij})\right]^{-1/2}\widehat{\sigma}_{e}\widehat{e}_{ijk}.$$ The specific steps of nonparametric bootstrap are described as follows:

For $m=1, ..., B$,
\begin{enumerate}
\item Sample $u_{ij}^{(m)c}$ with replacement from $\widehat{u}_{ij}^{c}$ and sample $e_{ijk}^{(m)c}$ with replacement from $\widehat{e}_{ijk}^{c}$.
\item Compute $y_{ijk}^{(m)}=x_{ij}'\gammavechat+u_{ij}^{(m)c}+e_{ijk}^{(m)c}$.
\item Fit the semi-parametric model to the bootstrapped sample $y_{ijk}^{(m)}$.
\end {enumerate}
The CI with confidence level $1-\alpha$ for a parameter of interest, $\theta$, is calculated by taking the lower and upper $\alpha/2$ quantiles of the bootstrap estimates. For a sequence of bootstrap estimates $\widehat{\theta}^{(1)}, \dots, \widehat{\theta}^{(B)} $, a bias-corrected CI, proposed by \citeN{EfronTibshirani1993}, can be computed by taking the $B\Phi(2z_{q}+z_{\alpha/2})$ and $B\Phi(2z_{q}+z_{1-\alpha/2})$ ordered values, where $q$ denotes the proportion of bootstrap values less than $\widehat{\theta}$, $\Phi(\cdot)$ is the cumulative distribution function and $z_{(\cdot)}$ is the quantile function of the standard normal distribution.

\section{Simulation Study} \label{sec: Simulation Study}
The objective of the simulation study is to investigate the performance of the proposed parameter estimation and inference procedures. We will examine the bias, standard derivation (SD), and mean square error (MSE) of the parameter estimators and the estimated baseline degradation function. We also will investigate the coverage probability (CP) of the bootstrap-based CI procedure in Section \ref{sec: Statistical Inference}. An additional simulation study will be conducted to investigate the performance of our semi-parametric model under model misspecification. 

\subsection{Performance of Parameter Estimators}
\subsubsection{Simulation Settings}
\begin{figure}
     \begin{center}
             \subfigure[Spline bases] {
        \includegraphics[width=0.47\textwidth]{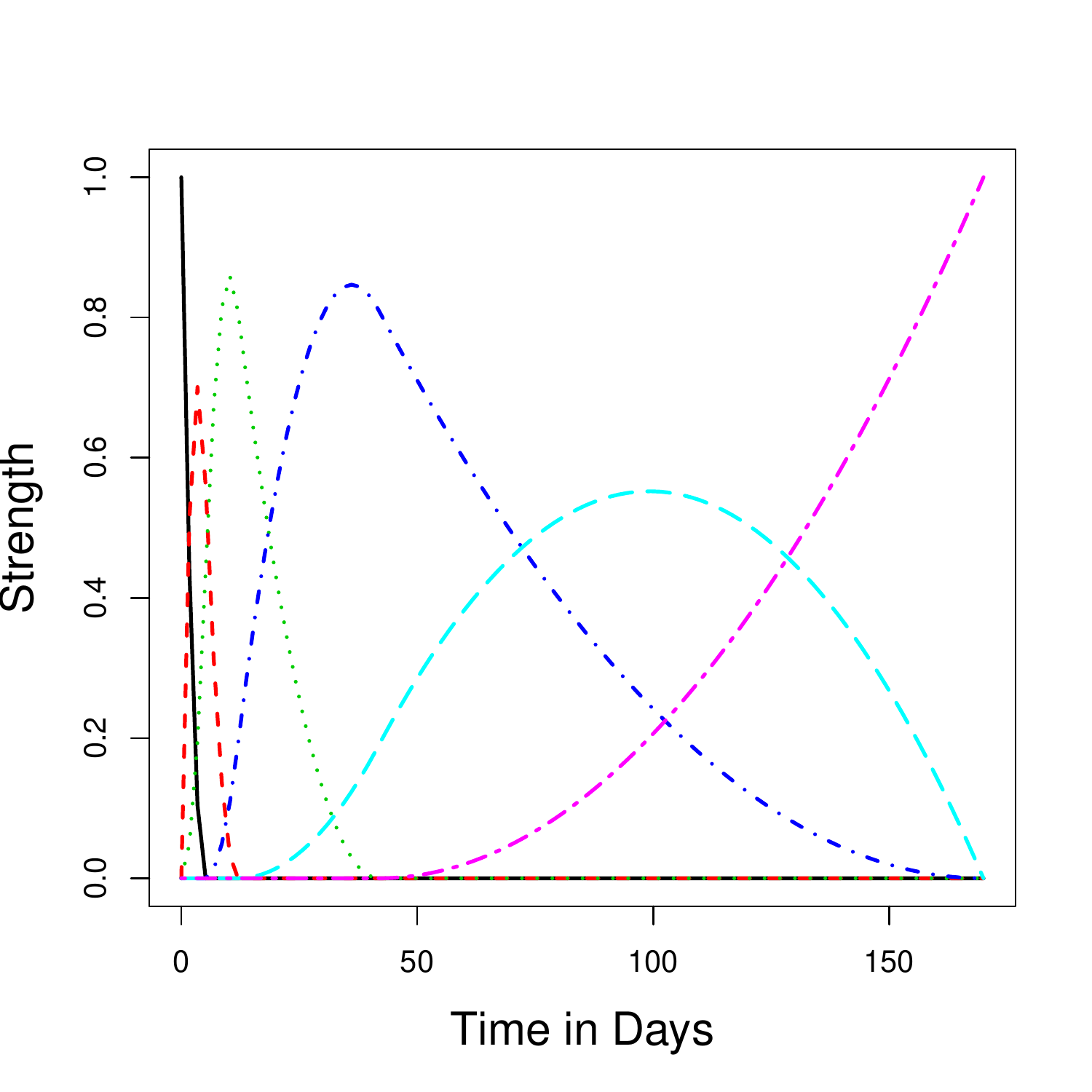}
                }
        \subfigure[Baseline degradation path] {
        \includegraphics[width=0.47\textwidth]{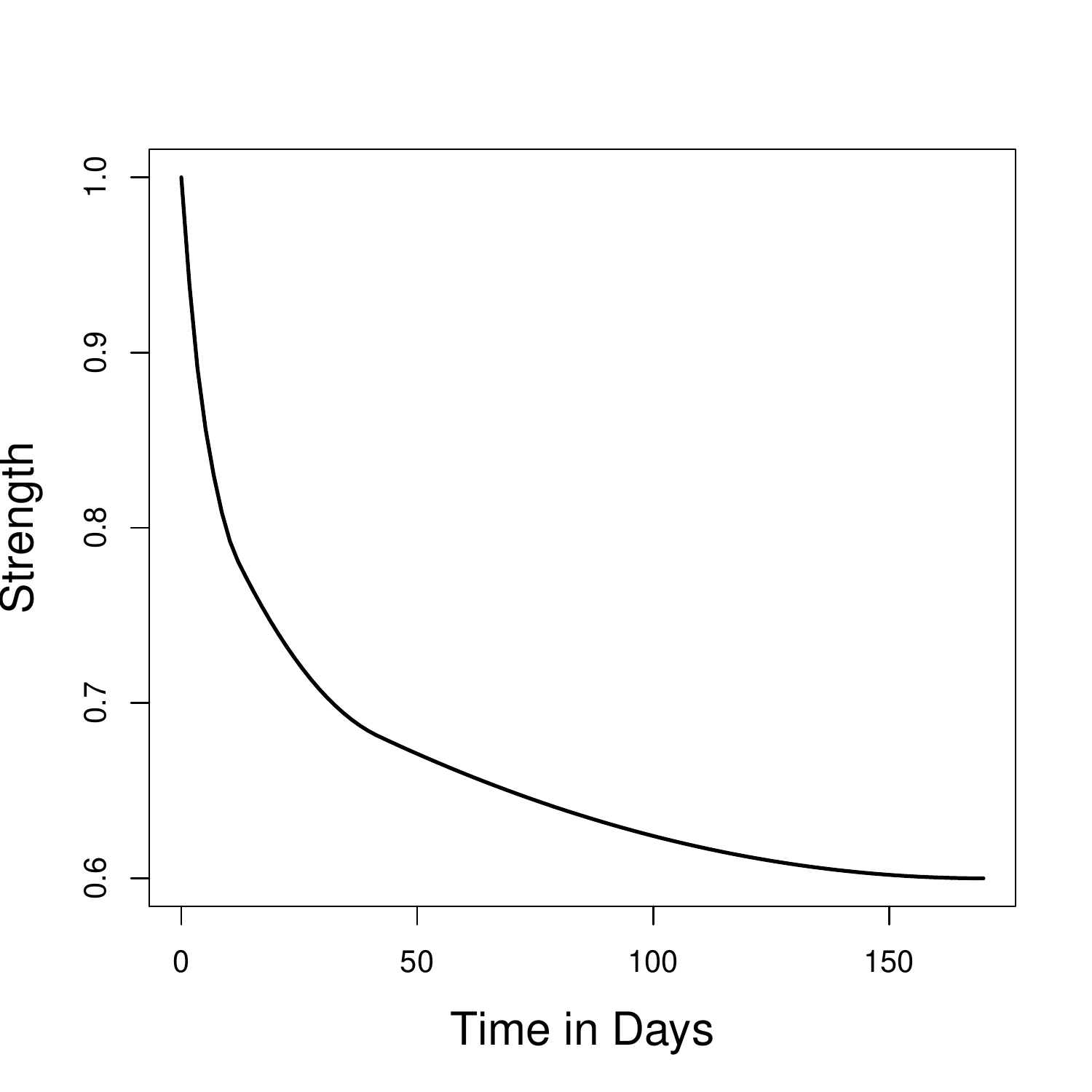}
        }
\caption{Spline bases and baseline degradation path used in simulation study.} \label{fig: Spline bases and baseline degradation curve used in simulation study}
 \end{center}
\end{figure}

We consider two different sets of $n=\{3,6\}$ temperature levels and three different sets of $J_n=\{5,10,15\}$ measuring times. The specific settings are summarized in Table \ref{tab:sim}. Ten samples are tested at each combination of temperature level and measuring times. The data are simulated from the following model:
\begin{gather} \label{eqn: true semi-parametric model in simulation study}
y_{ijk}=\sum_{l=1}^{p} r_{l} B_{q, l}[\eta_{i}(t_{ij};\beta)] + \varepsilon_{ijk},
\end{gather}
where the degree of the B-splines is $q=2$, and number of interior knots is $N=3$. The knot locations are the sample quantiles. Figure \ref{fig: Spline bases and baseline degradation curve used in simulation study} gives the spline basis functions and the baseline degradation function for scenario $n=3$,  $J_{n}=5$. The true parameters in the model are $\beta=0.83, \gammavec=(1, 0.9, 0.8, 0.7, 0.6, 0.6)'$,  and $(\sigma, \rho)'=( 0.019, 0.2)'$.

For each scenario, 500 datasets are generated and the bias, SD, and MSE of the parameter estimators and baseline degradation curves are calculated. The quantile and bias-corrected CI are computed based on $B=1,000$ bootstrap samples and the CP is also computed.

\begin{table}
\begin{center}
\caption{Selected temperature levels and time points for the simulation studies.}\label{tab:sim}
\begin{tabular}{ c|c|l}\hline\hline
Settings & Number of Temp. Levels ($n$) & Temperature Levels ($\degreeC$)\\\hline
Temperature setting 1 &   3 & 50, 65, 80          \\\hline
Temperature setting 2 &   6 & 30, 40, 50, 60, 70, 80\\\hline\hline
        & Number of Time Points ($J_{n}$) & Measuring Times (Hours)\\\hline
Time point setting 1 &  5 & 8, 25, 75, 130, 170 \\ \hline
\multirow{2}{*}{Time point setting 2} & \multirow{2}{*}{10} & 5, 10,  30,  50,  70,  90, 110,\\
                     &    &130, 150, 170\\ \hline
\multirow{3}{*}{Time point setting 3} & \multirow{3}{*}{15} & 10,  30, 40,  50,  60,  70,  80,\\
                     &    & 90, 100, 110, 120, 130,\\
                     &    & 140, 150, 170\\\hline\hline
\end{tabular}
\end{center}
\end{table}

\subsubsection{Simulation Results}

\begin{figure}
     \begin{center}

             \subfigure[$\beta$] {
        \includegraphics[width=0.3\textwidth]{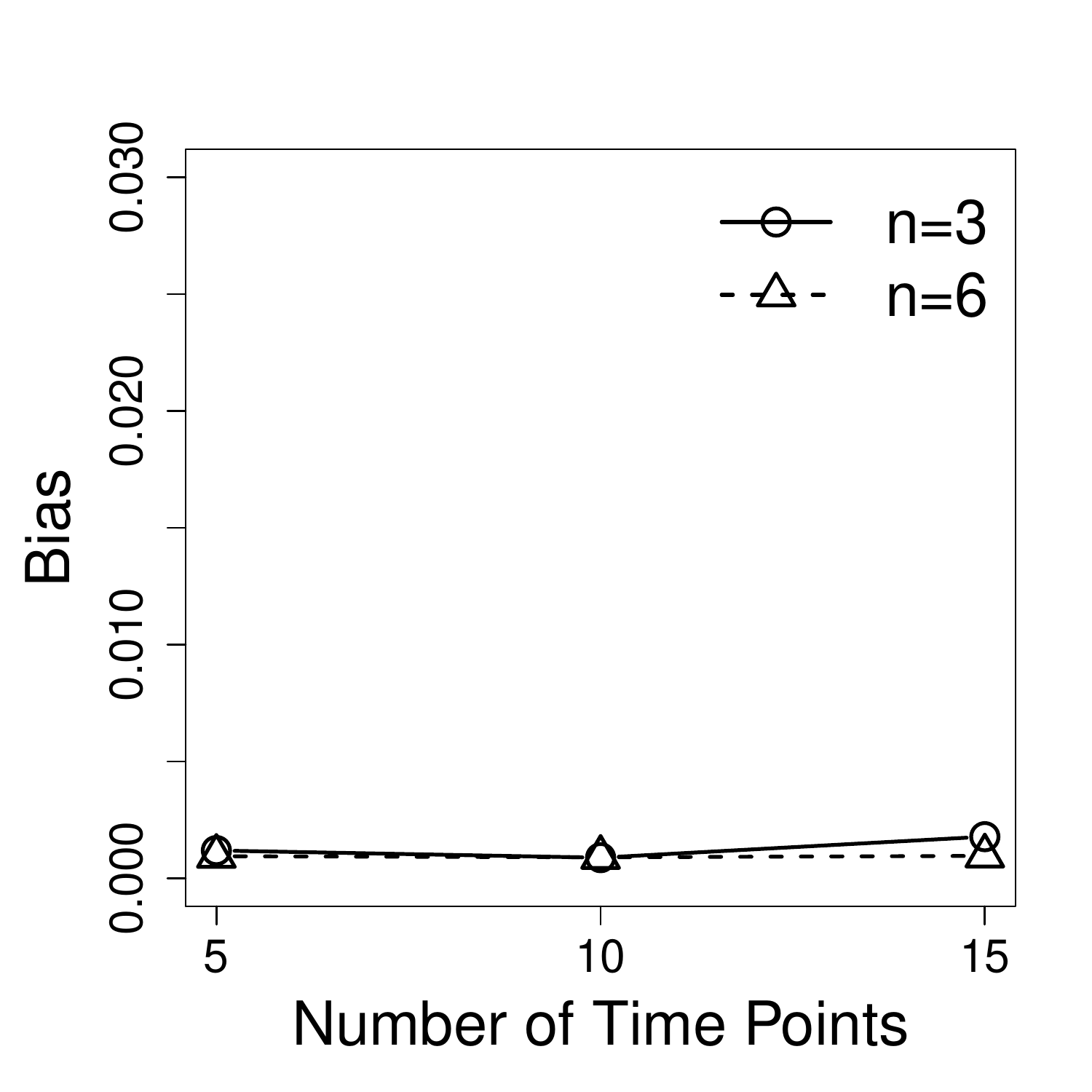}
                }
        \subfigure[$\sigma$] {
        \includegraphics[width=0.3\textwidth]{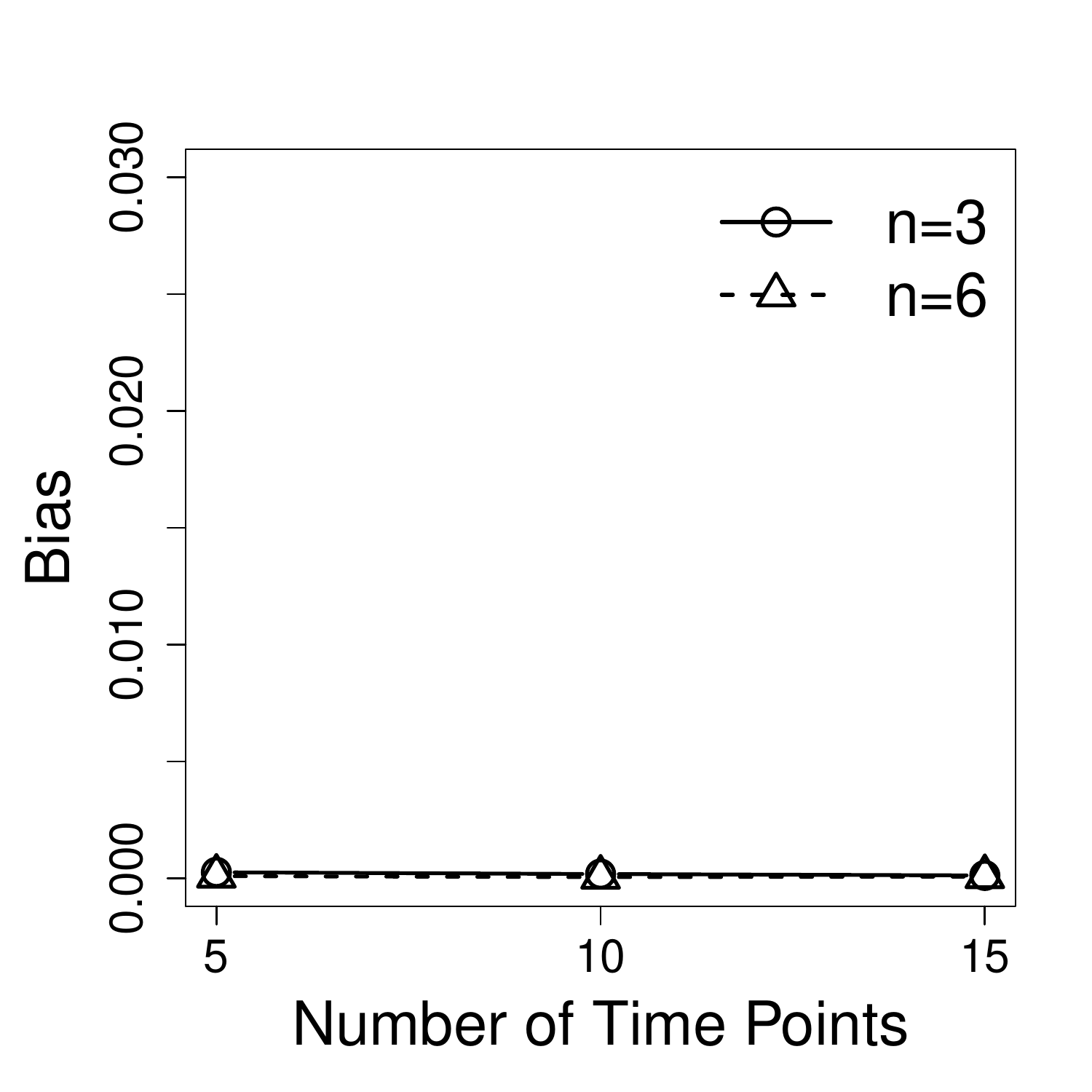}
        }
        \subfigure[$\rho$] {
        \includegraphics[width=0.3\textwidth]{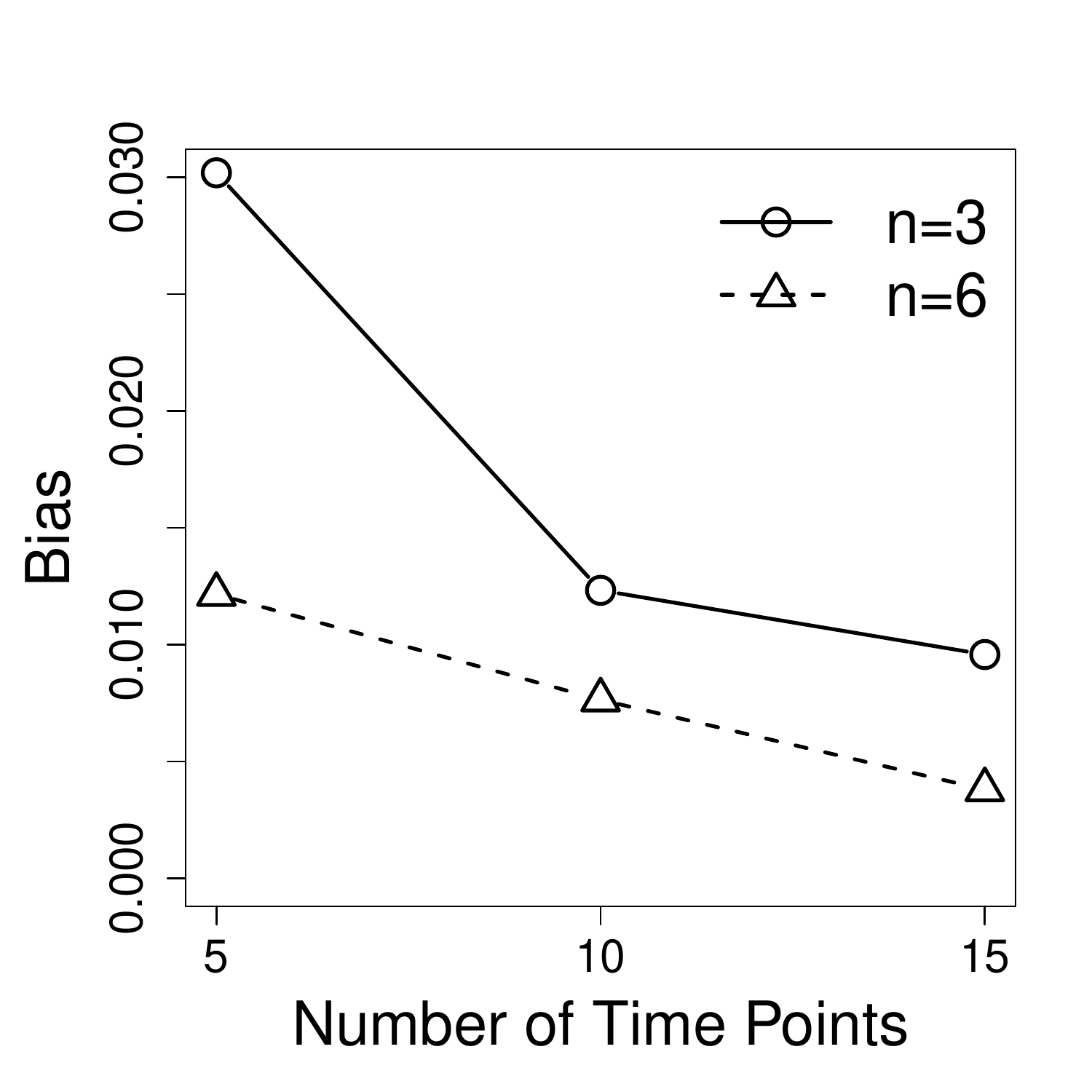}
        }     \\
             \subfigure[$\beta$] {
        \includegraphics[width=0.3\textwidth]{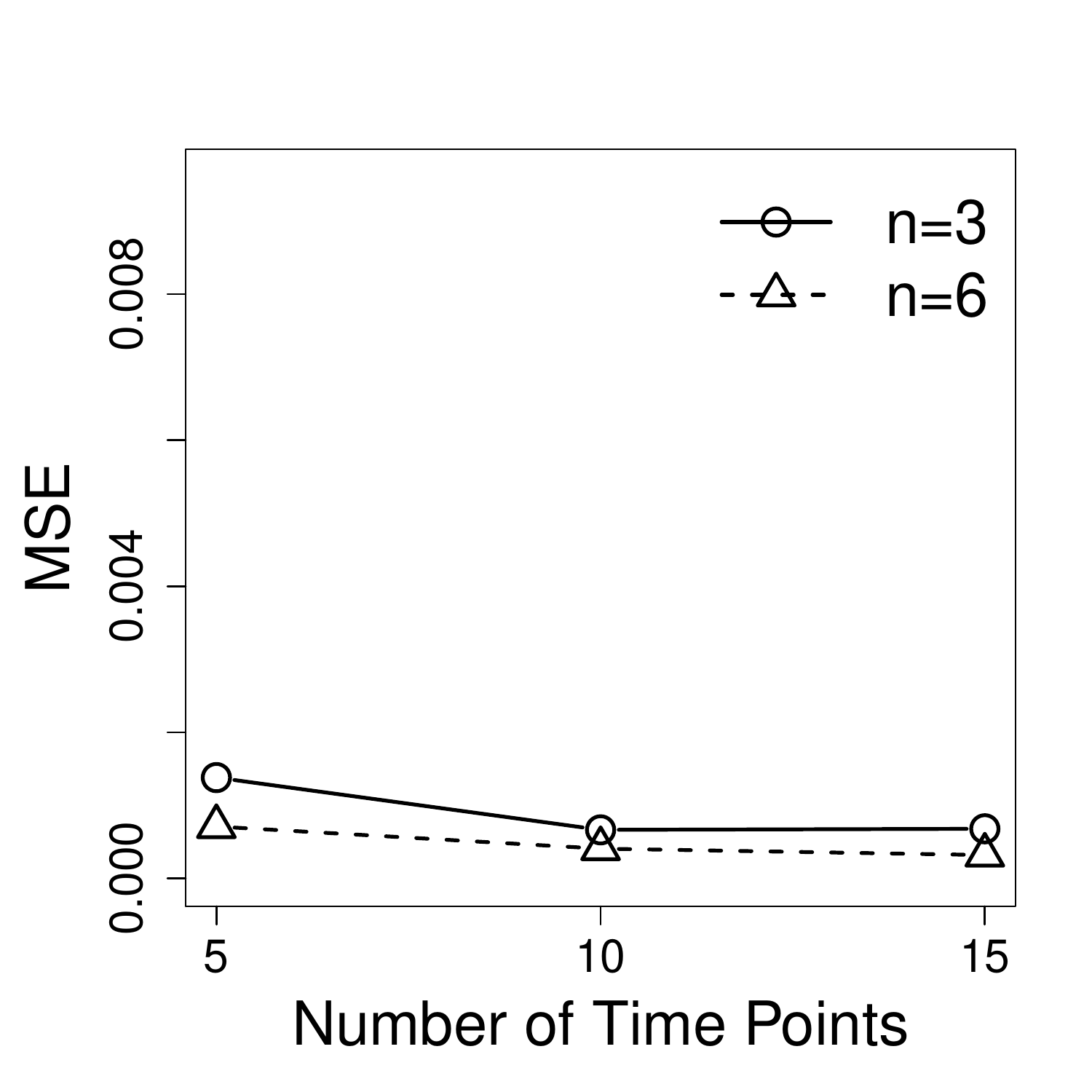}
                }
        \subfigure[$\sigma$] {
        \includegraphics[width=0.3\textwidth]{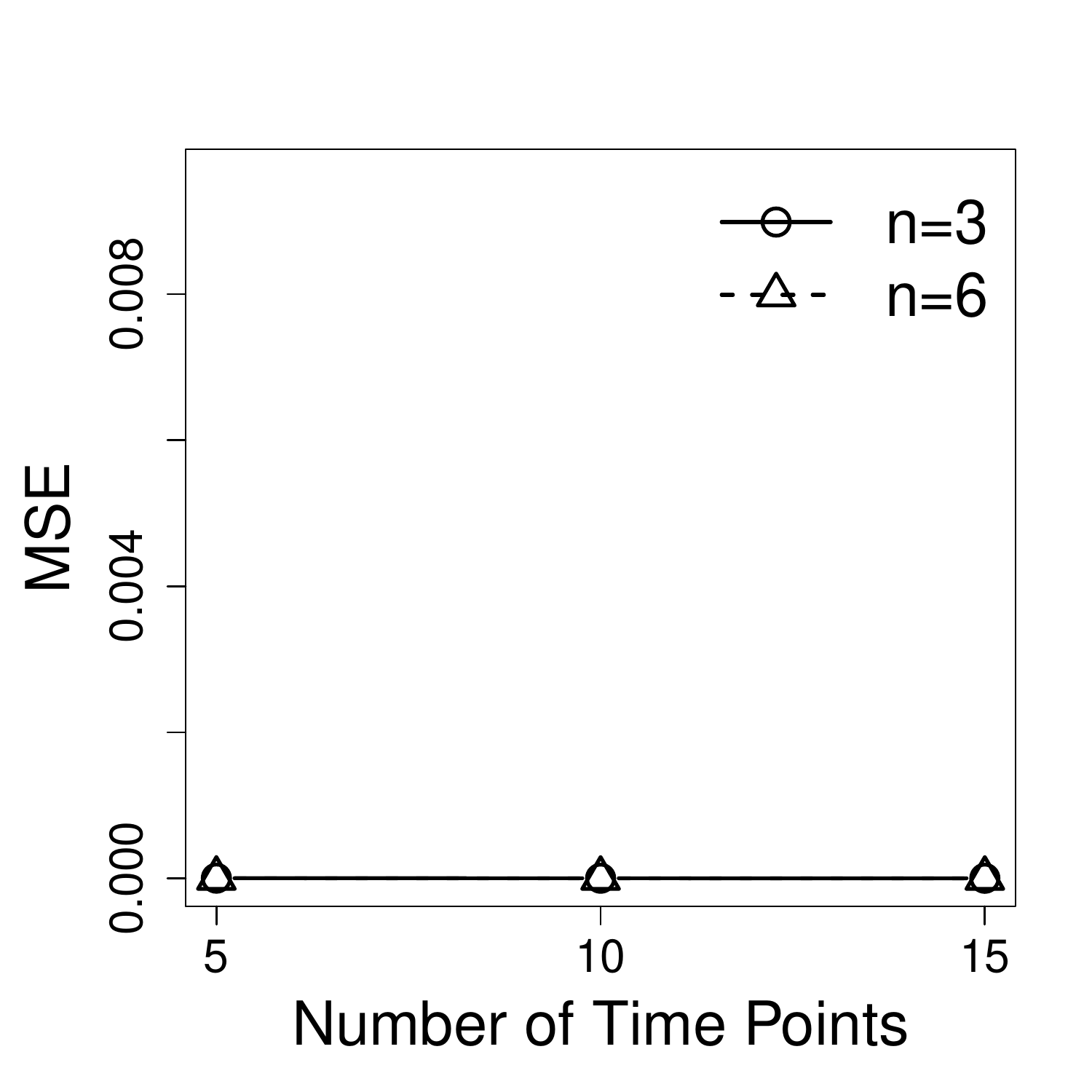}
        }
        \subfigure[$\rho$] {
        \includegraphics[width=0.3\textwidth]{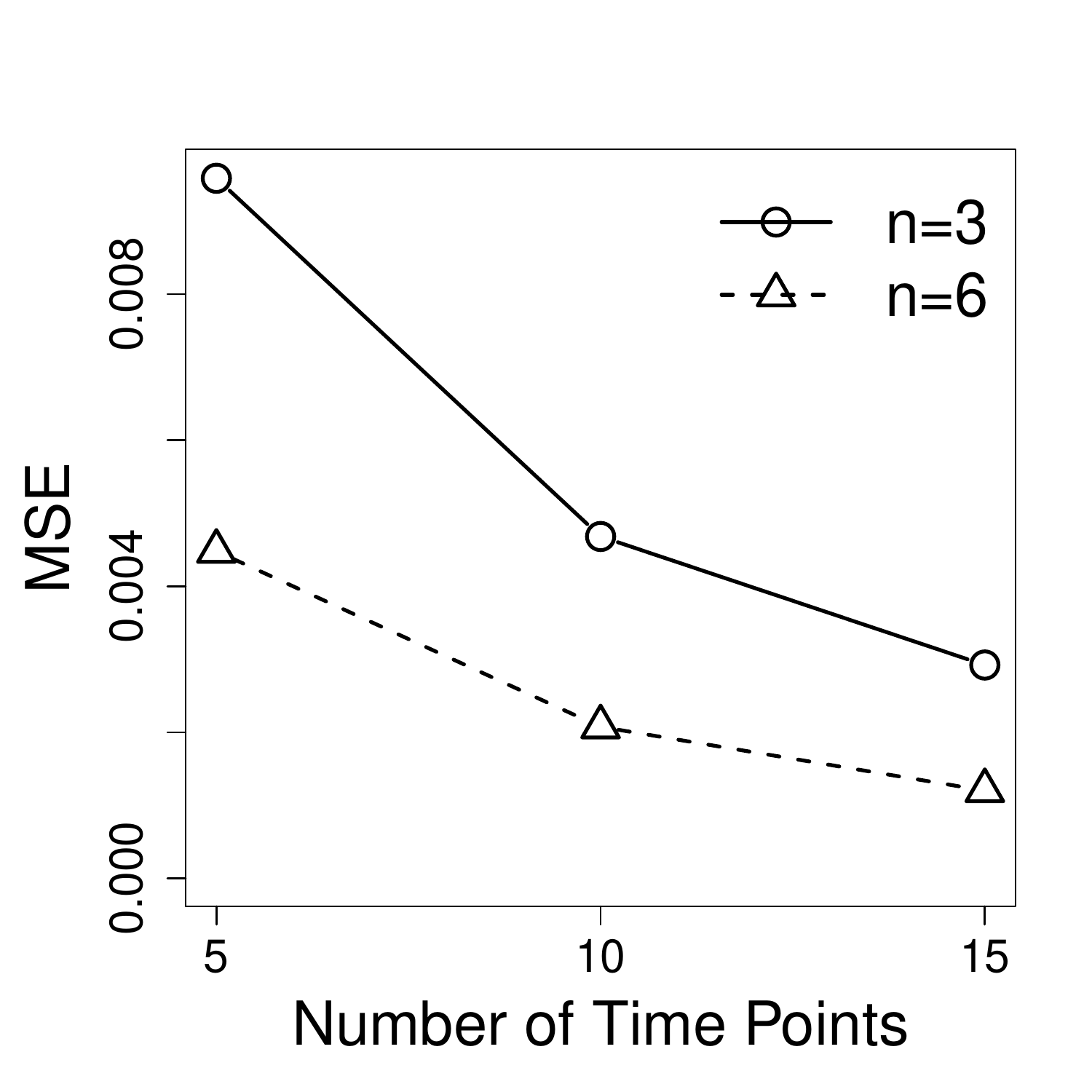}
        }
              \caption{Empirical bias and MSE of parameter estimators for $(\beta, \sigma, \rho)'$.} \label{fig: Bias and MSE of parameters}
 \end{center}
\end{figure}

\begin{figure}
\begin{center}{
\includegraphics[width=0.55\textwidth]{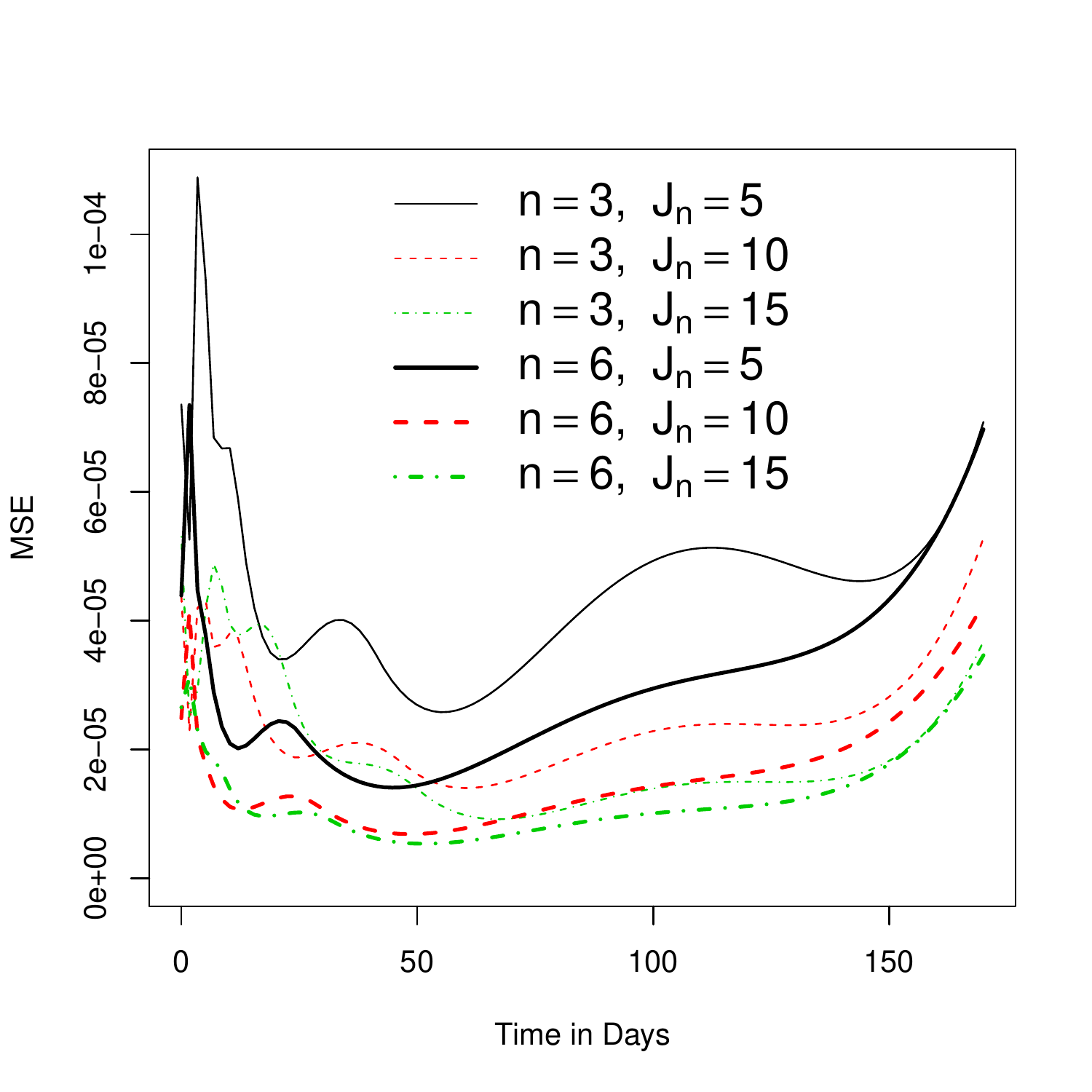}
}
\end{center}
\caption{Empirical pointwise MSE for the estimator of the baseline degradation path.}
\label{fig: Pointwise MSE curve for baseline degradation curve.}
\end{figure}

Figure \ref{fig: Bias and MSE of parameters} shows the bias and MSE of parameter estimators. Figure \ref{fig: Pointwise MSE curve for baseline degradation curve.} shows the pointwise MSE curves of baseline degradation curves. We found out that MSE of point estimators and baseline degradation curves decrease as either number of temperature levels or time points increases. Even when the number of temperature levels and time points are both small, biases of $\beta$ and $\sigma$ are small, while bias of $\rho$ is large. However, when either number of temperature levels or time points is large, the estimates of $\beta, \sigma$ and $\rho$ are all close to the true values.

\begin{figure}
     \begin{center}

             \subfigure[$\beta$, Quantile-based CI] {
        \includegraphics[width=0.3\textwidth]{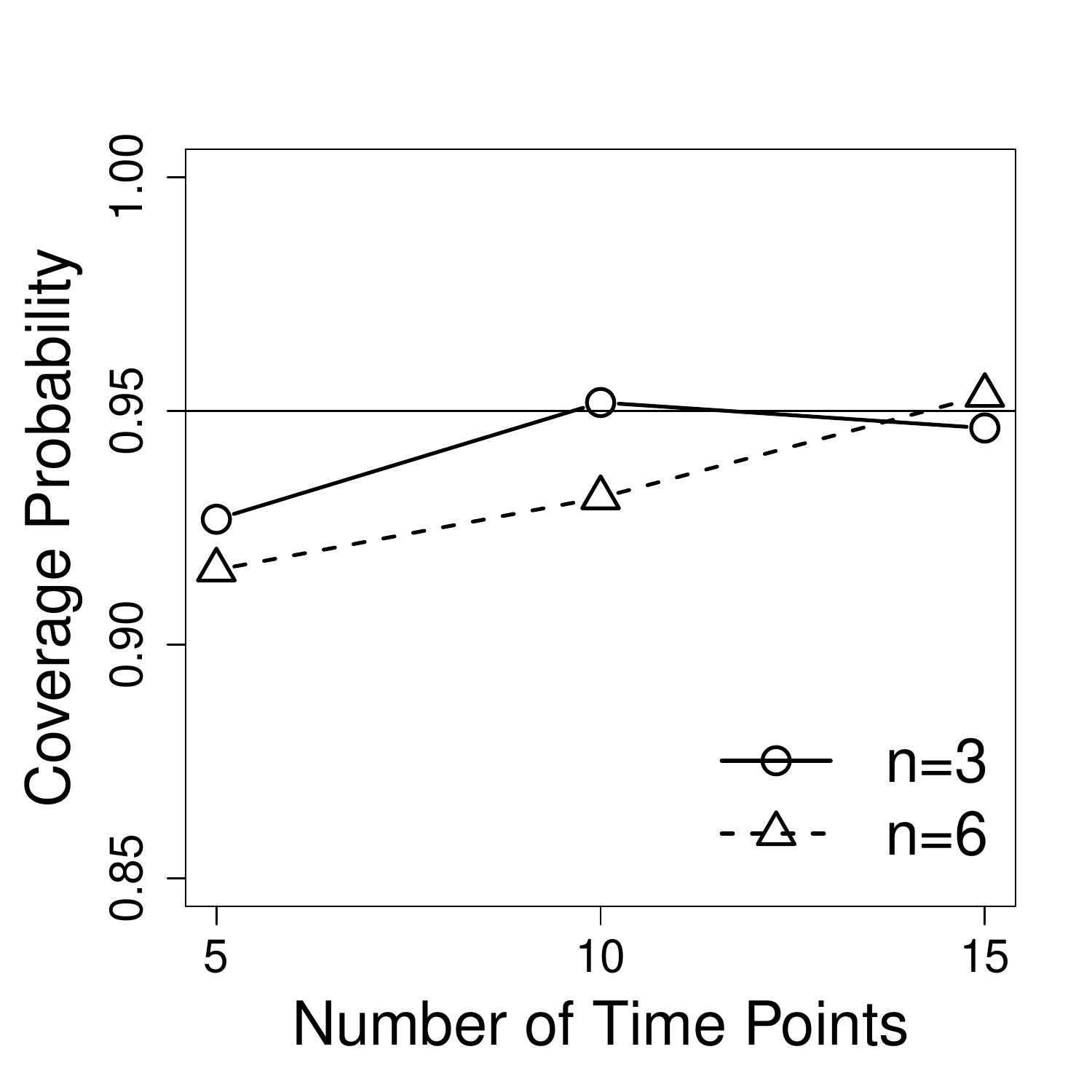}
                }
        \subfigure[$\sigma$, Quantile-based CI] {
        \includegraphics[width=0.3\textwidth]{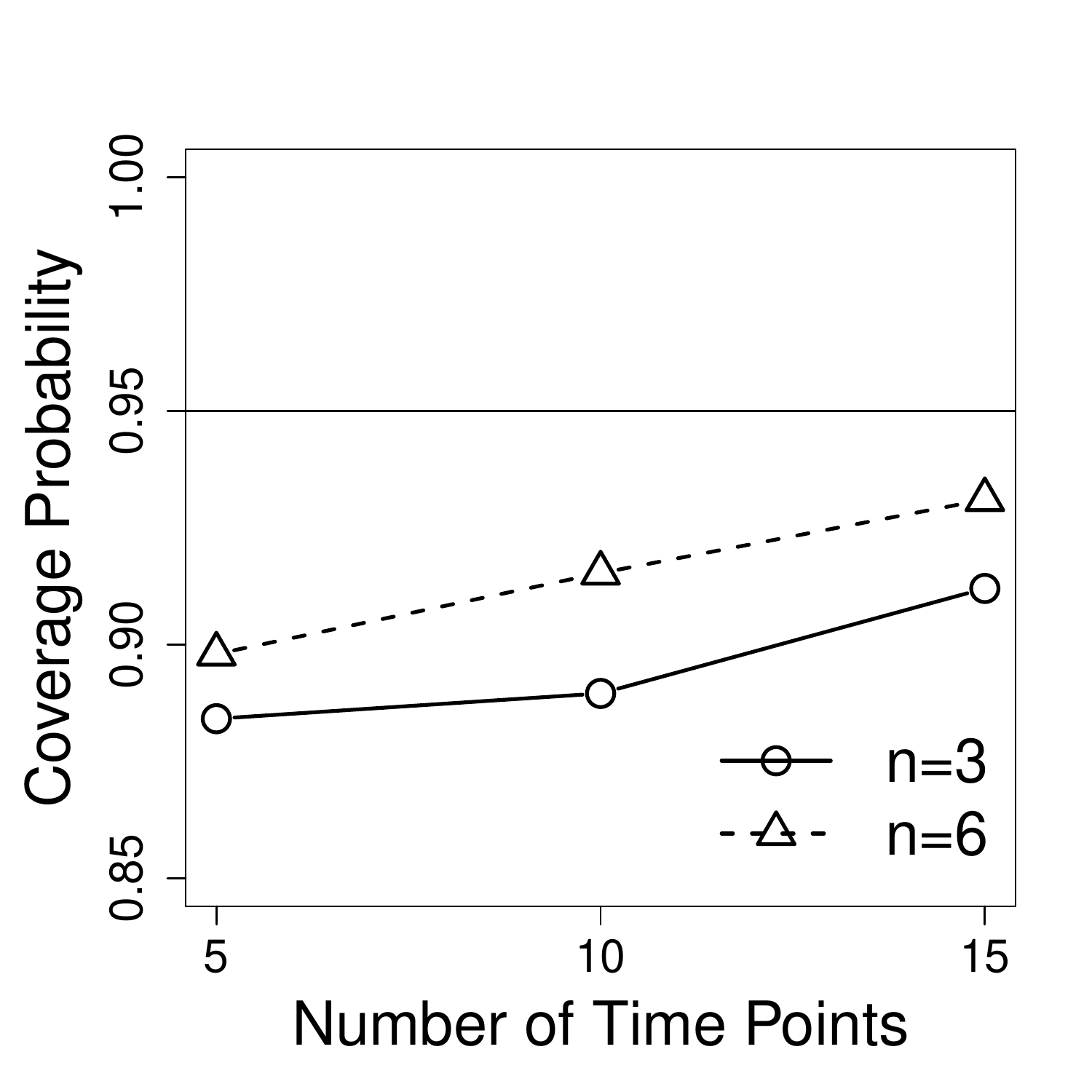}
        }
        \subfigure[$\rho$, Quantile-based CI] {
        \includegraphics[width=0.3\textwidth]{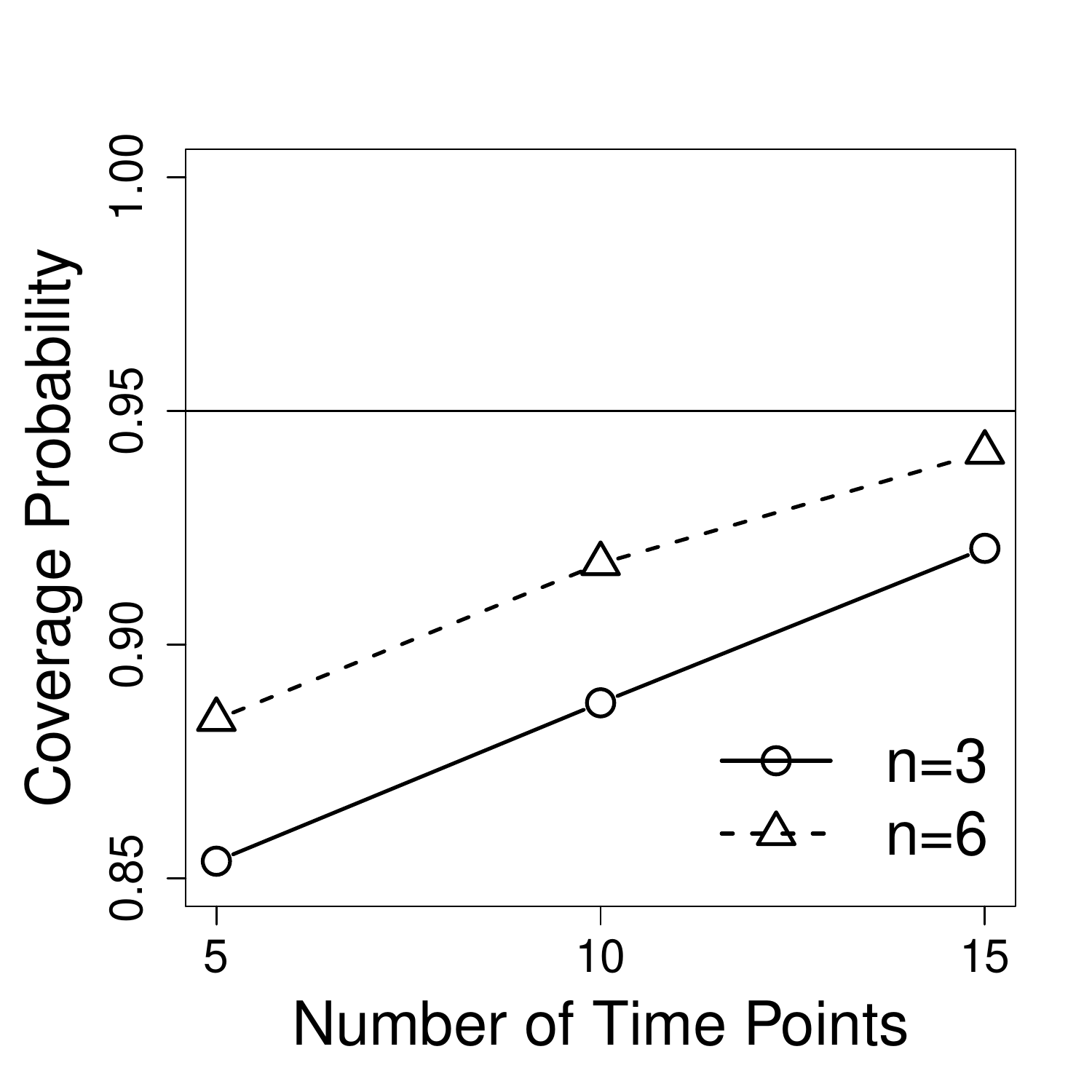}
        }     \\
             \subfigure[$\beta$, Bias-corrected CI] {
        \includegraphics[width=0.3\textwidth]{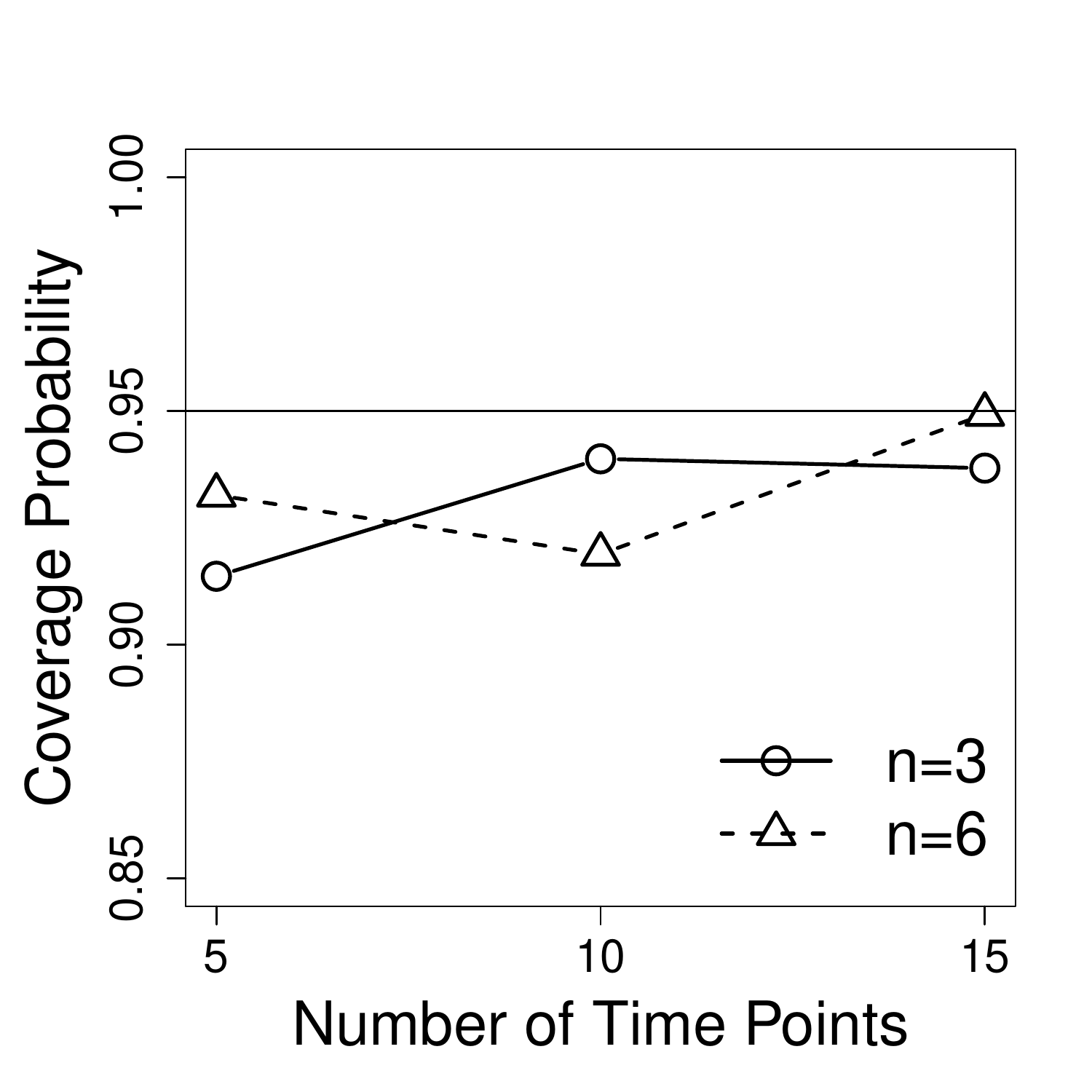}
                }
        \subfigure[$\sigma$, Bias-corrected CI] {
        \includegraphics[width=0.3\textwidth]{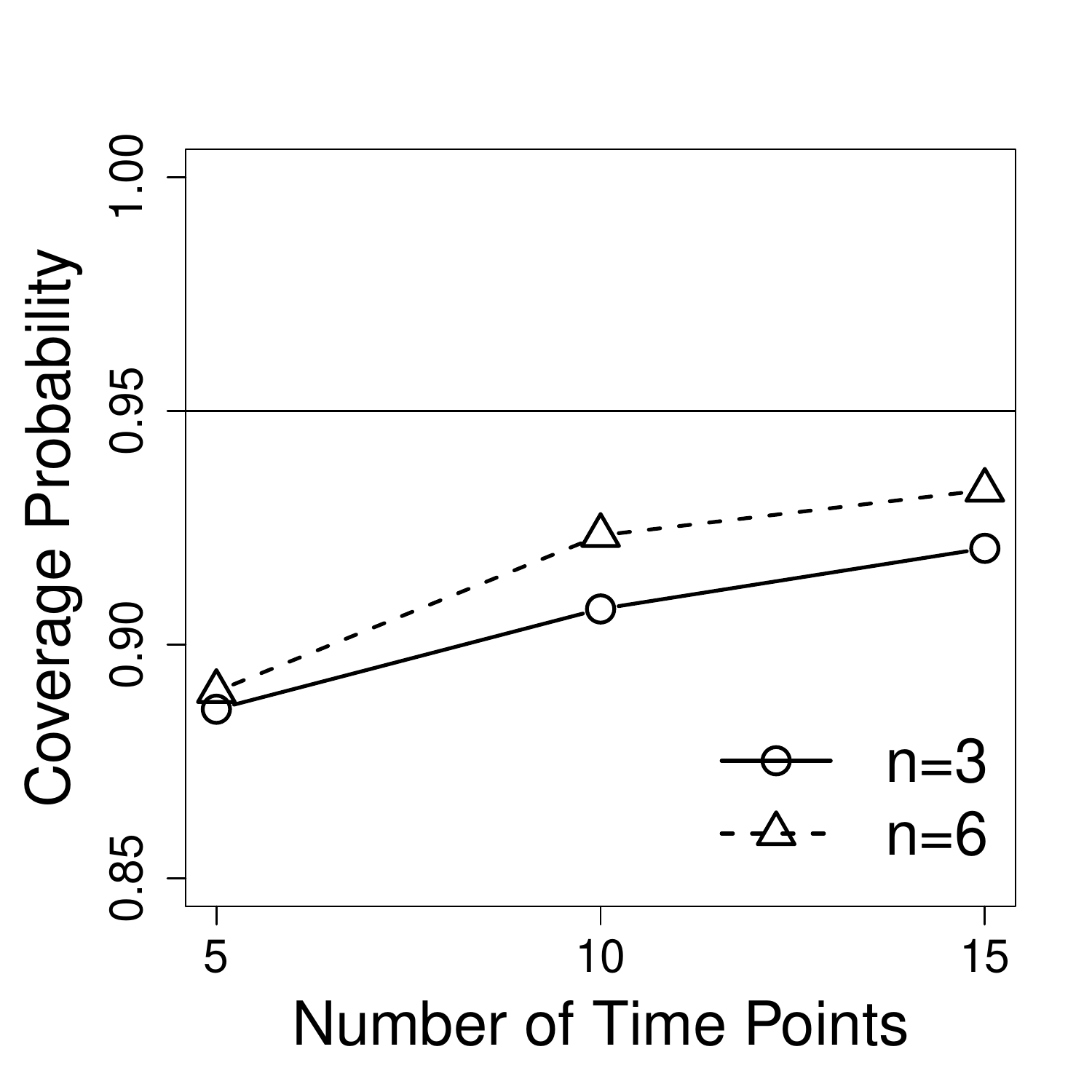}
        }
        \subfigure[$\rho$, Bias-corrected CI] {
        \includegraphics[width=0.3\textwidth]{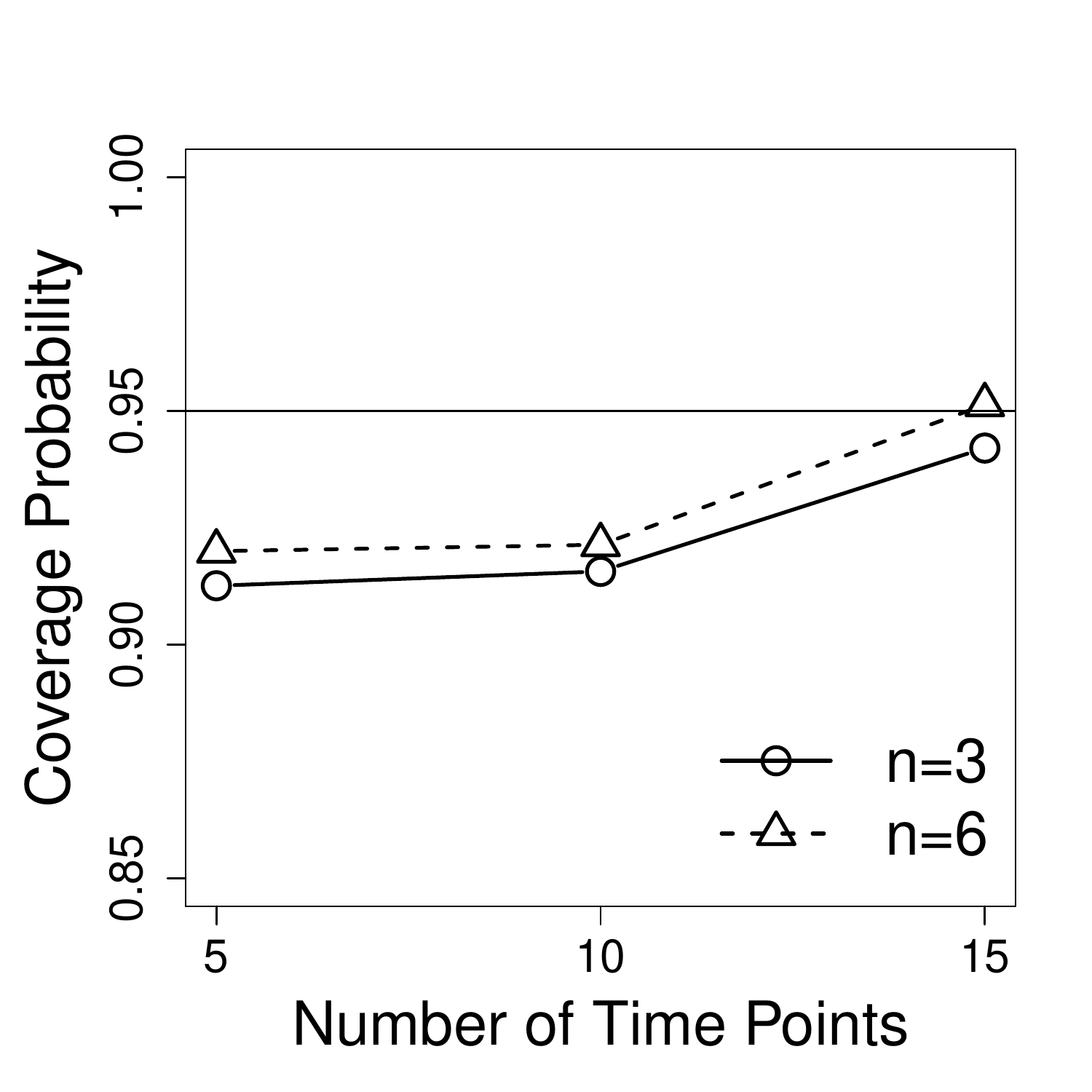}
        }
              \caption{CP of the CI procedures for parameters $(\beta, \sigma, \rho)'$, using quantile-based and bias-corrected methods, respectively.} \label{fig: CPs of CI of parameter estimators.}
 \end{center}
\end{figure}

\begin{figure}
     \begin{center}
             \subfigure[Quantile-based CI] {
        \includegraphics[width=0.47\textwidth]{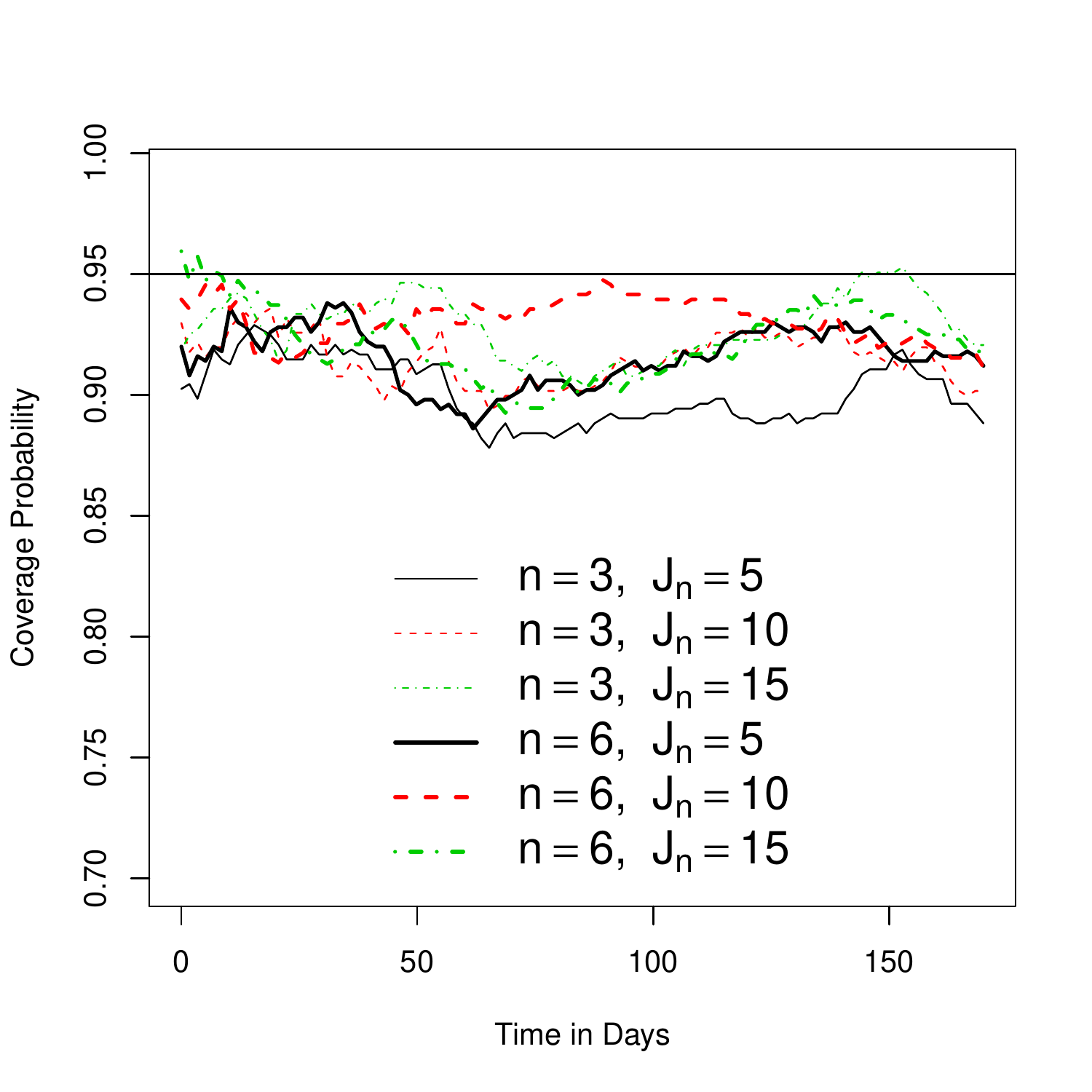}
                }
        \subfigure[Bias-corrected CI] {
        \includegraphics[width=0.47\textwidth]{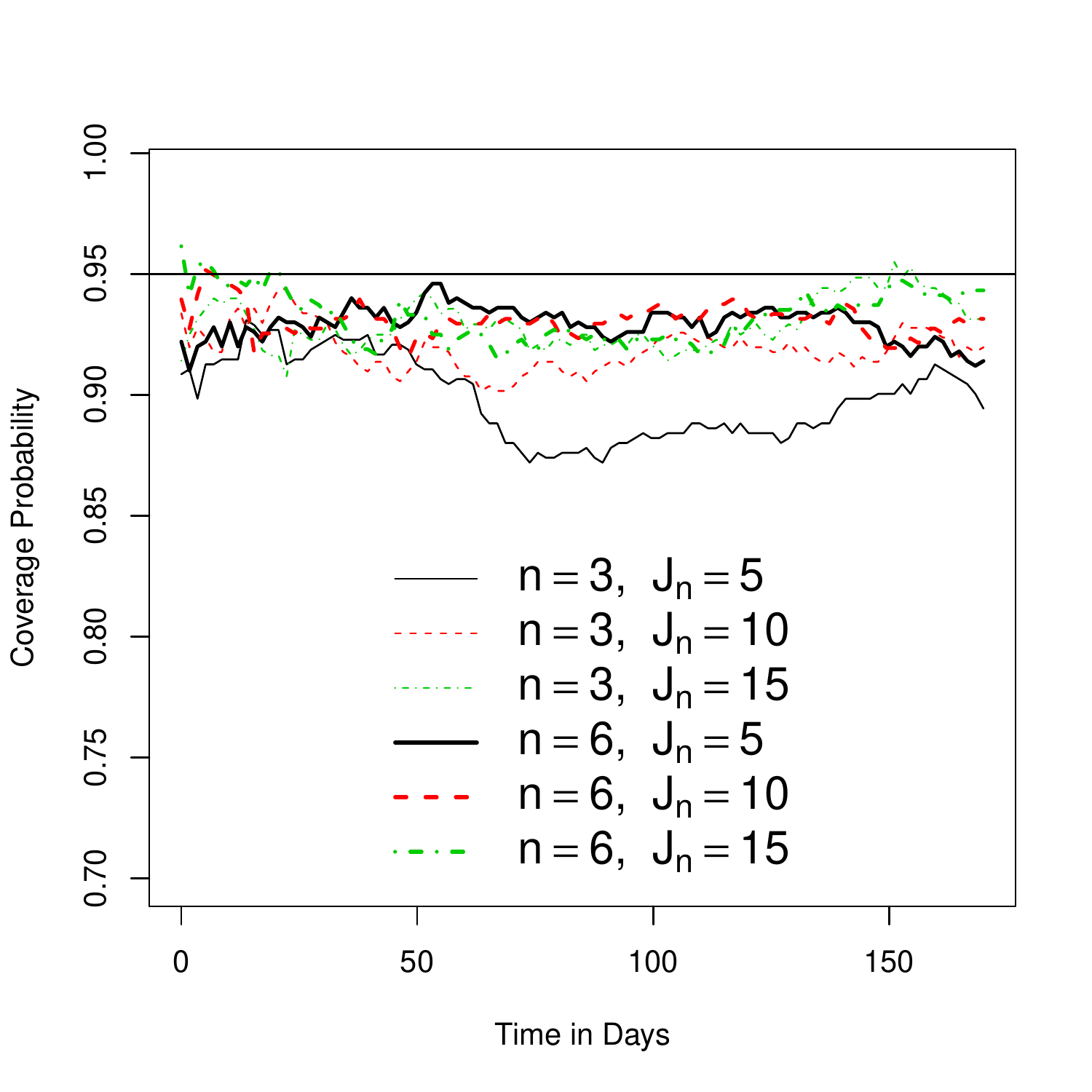}
        }
\caption{Pointwise CP of the CI procedure for baseline degradation path, using quantile-based and bias-corrected methods, respectively.}  \label{fig: CPs of pointwise CI of baseline degradation curve.}
 \end{center}
\end{figure}

Figures \ref{fig: CPs of CI of parameter estimators.} and \ref{fig: CPs of pointwise CI of baseline degradation curve.} present the CP for quantile-based CI and bias-corrected CI of the parameter estimators and baseline degradation curves. The performance of bias-corrected CI seems to be similar for $\beta$, and better for $\sigma$, $\rho$ and baseline degradation curve compared to quantile-based CI. For the parameter estimators, the CP of bias-corrected CI of $\beta$ is good when $n$ or $J_{n}$ is small. However, the CP of bias-corrected CI of $(\sigma, \rho)'$ are overall slightly less than the desired confidence level. For the baseline degradation function,  the CP of pointwise bias-corrected CI are poor when $n=3$ and $J_{n}=5$. The performance of pointwise bias-corrected CI improve as $n$ and $J_{n}$ increases. Overall, the results show that the performance of the estimation and inference procedures are good.

\subsection{Performance under Model Misspecification} \label{sec: Comparison}
\subsubsection{Simulation Settings}
\begin{figure}
\begin{center}{
\includegraphics[width=0.55\textwidth]{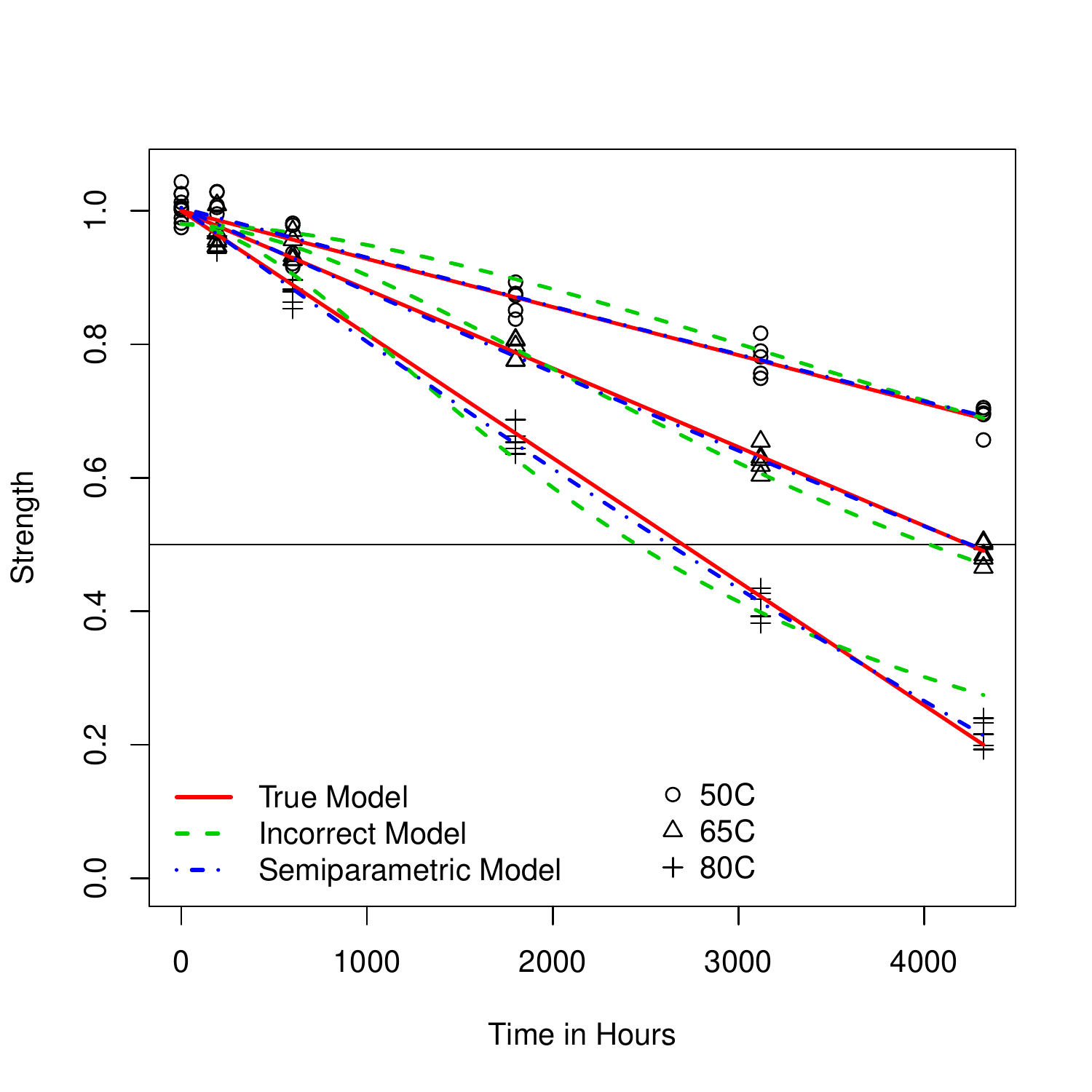}
}
\end{center}
\caption{Plot of simulated data and fitted degradation paths based on the true and incorrect parametric models, and semi-parametric model.}
\label{fig: Scatter plot of simulated data}
\end{figure}

In this simulation study, the data are simulated according to a parametric model, but the semi-parametric model is fit to the data. The temperature levels are set at $50\degreeC$, $65\degreeC$, $80\degreeC$ and the measuring times are set at $192, 600 ,1800, 3120$, and  $4320$ hours. There are $10$ measurements at time $0$ and $5$ measurements at all other measuring times. The data are simulated from the model
\begin{gather} \label{eqn: true parametric model in simulation study}
y_{ijk}=\beta_{0}+\beta_{1}\exp(\beta_{2} x_{i} ) \tau_{j} + \varepsilon_{ijk},
\end{gather}
where $\tau_{j}=\textrm{Hour}_{j}, x_{i}=-11605/(\textrm{Temp}_{i}+273.15)$. The true parameters are $\betavec=(\beta_{0}, \beta_{1}, \beta_{2})'=(1, -3.5, 0.3)'$, and $(\sigma, \rho)'=( 0.02, 0)'$. It is rare for the true model to be known exactly, so we also consider the case when a different parametric model from the true one is fit to the data. The incorrect parametric model, adapted from \shortciteN{VacaTrigoMeeker2009}, is given by
\begin{gather} \label{eqn: Incorrect parametric model in the simulation study}
y_{ijk} = \frac{\alpha}{1+\left[ \frac{t_{ij}}{\exp(\beta_{0}+\beta_{1} x_{i})}\right]^{\gamma}}+ \varepsilon_{ijk},
\end{gather}
with parameters $(\alpha, \beta_0, \beta_1, \gamma)'$ in the mean structure. We fit the true model (\ref{eqn: true parametric model in simulation study}), the incorrect parametric model (\ref{eqn: Incorrect parametric model in the simulation study}), and our semi-parametric model (\ref{eqn: semi-parametric degradation model}) to the simulated data. Figure \ref{fig: Scatter plot of simulated data} shows one case of the simulated data and the fitted degradation paths.


\subsubsection{Simulation Results}
To assess the fit of our semi-parametric model, we compare the fitted degradation path to the true degradation path using the integrated mean square error (IMSE) of the baseline degradation function, which is defined as
\begin{align*}
\textrm{IMSE}&= \int_{0}^{t_{m}} \E \left\{ \left[\widehat{g}(t; \gammavec)- g(t; \gammavec)\right]^{2} \right\}dt \\
&=  \int_{0}^{t_{m}}  \left\{ \E  \left[\widehat{g}(t; \gammavec)  \right]- g(t; \gammavec) \right\}^{2}dt +  \int_{0}^{t_{m}}  \Var  \left[\widehat{g}(t; \gammavec) \right]dt =\textrm{IBias}^2+\textrm{IVar},
\end{align*}
where $t_{m}$ is the maximum time under the maximum level of the accelerating variable. As there is no closed-form expressions for IMSE, IBias and IVar, we report the empirical results. Table \ref{tab: EISBias, EIVar, EIMSE} presents these results, which indicate that the performance of our semi-parametric model is good. The largest contribution to the root IMSE comes from the variance component. Thus, it is not surprising that the incorrect parametric model (\ref{eqn: Incorrect parametric model in the simulation study}) performs the worst in capturing the true degradation path.

For each simulated dataset, the MTTF at $30\degreeC$ is calculated based on the true parametric model (\ref{eqn: true parametric model in simulation study}), incorrect parametric model (\ref{eqn: Incorrect parametric model in the simulation study}) and the semi-parametric model. The mean, bias, standard derivation and root MSE of the MTTF for each of the different models based on $600$ datasets are summarized in Table \ref{tab: Estimated mean, bias, standard derivation, and MSE of the MTTF estimators}. The results indicate that the estimate of MTTF from our semi-parametric model is close to the true values, but with larger variance. The estimated MTTF from the incorrect parametric model (\ref{eqn: Incorrect parametric model in the simulation study}) has the largest bias. The results indicate our semi-parametric model performs quite well.

\begin{table}
\centering
\caption{Empirical IBias, root IVar (RIVar), and root IMSE (RIMSE) for the true model (\ref{eqn: true parametric model in simulation study}), incorrect model (\ref{eqn: Incorrect parametric model in the simulation study}), and the semi-parametric model.}
\label{tab: EISBias, EIVar, EIMSE}
\begin{tabular}{crrr}\hline\hline
Models & IBias & RIVar & RIMSE \\\hline
True Model              & 0.0003 & 0.0043 & 0.0043\\
  Incorrect Model           & 0.0267 & 0.0060 & 0.0274 \\
  Semi-parametric Model & 0.0003 & 0.0091 & 0.0091 \\\hline\hline
\end{tabular}
\end{table}

\begin{table}
\centering
\caption{Empirical mean, bias, SD, and root MSE (RMSE) of the MTTF estimators based on the true model (\ref{eqn: true parametric model in simulation study}), incorrect model (\ref{eqn: Incorrect parametric model in the simulation study}), and the semi-parametric model.} \label{tab: Estimated mean, bias, standard derivation, and MSE of the MTTF estimators}
\begin{tabular}{crrrr}\hline\hline
Models                 & Mean    & Bias   &   SD & RMSE \\\hline
True Model             &82.60 & 0.01 & 2.99 & 2.99 \\
Incorrect Model            &  85.82 & 3.20 & 3.75 & 4.93 \\
Semi-parametric Model  & 82.77 & 0.16 & 4.22 & 4.22\\\hline\hline
\end{tabular}
\end{table}

\section{Applications} \label{sec: Applications}
To help motivate the use of our semi-parametric model, we selected three published datasets from well-known examples of ADDT. The data for each example are summarized below.

\subsection{ADDT Datasets and Parametric Models}
\subsubsection{Adhesive Bond B Data} \label{sec: Adhesive Bond B Data}
\begin{figure}
     \begin{center}
             \subfigure[Parametric model] {
        \includegraphics[width=0.47\textwidth]{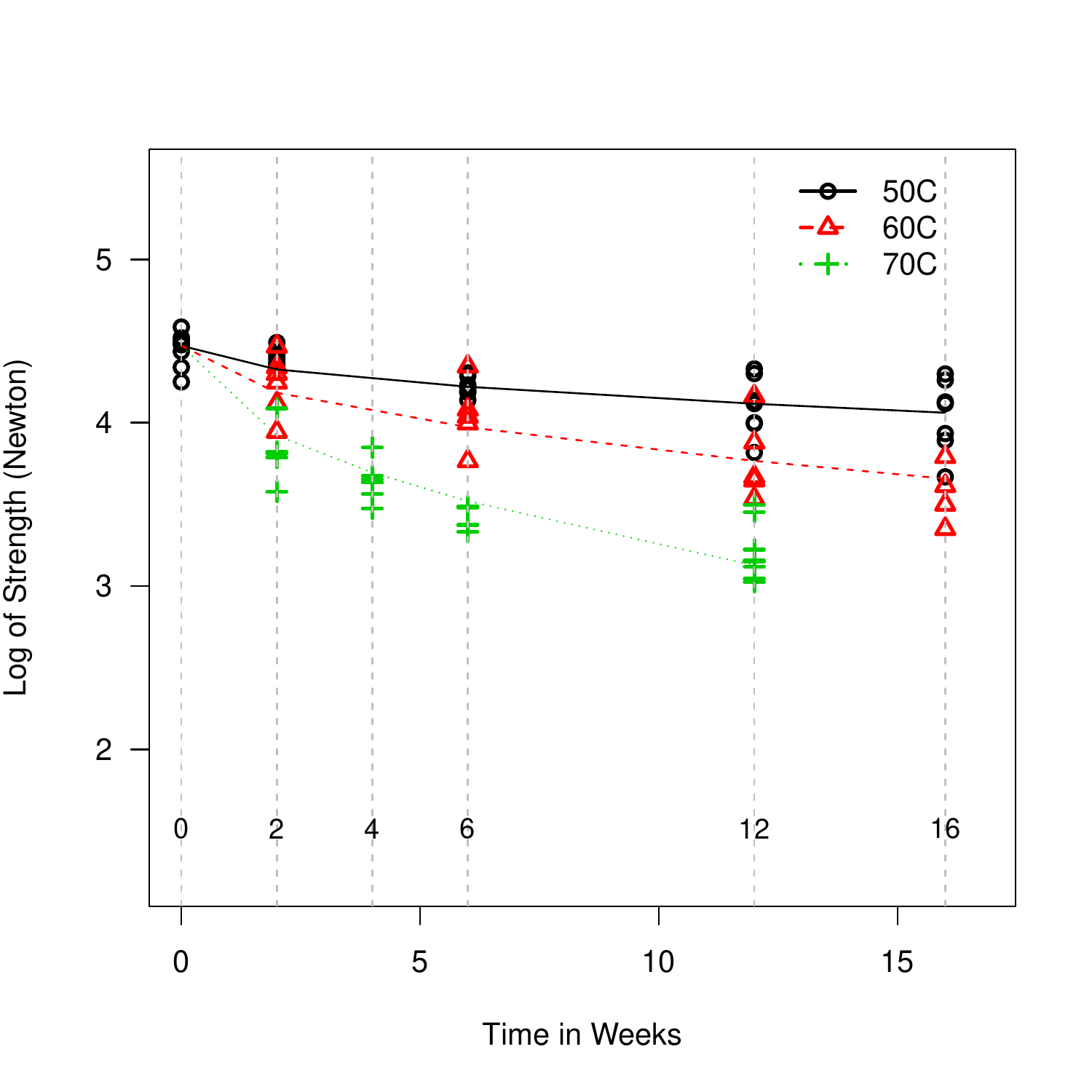}
                }
        \subfigure[Semi-parametric model] {
        \includegraphics[width=0.47\textwidth]{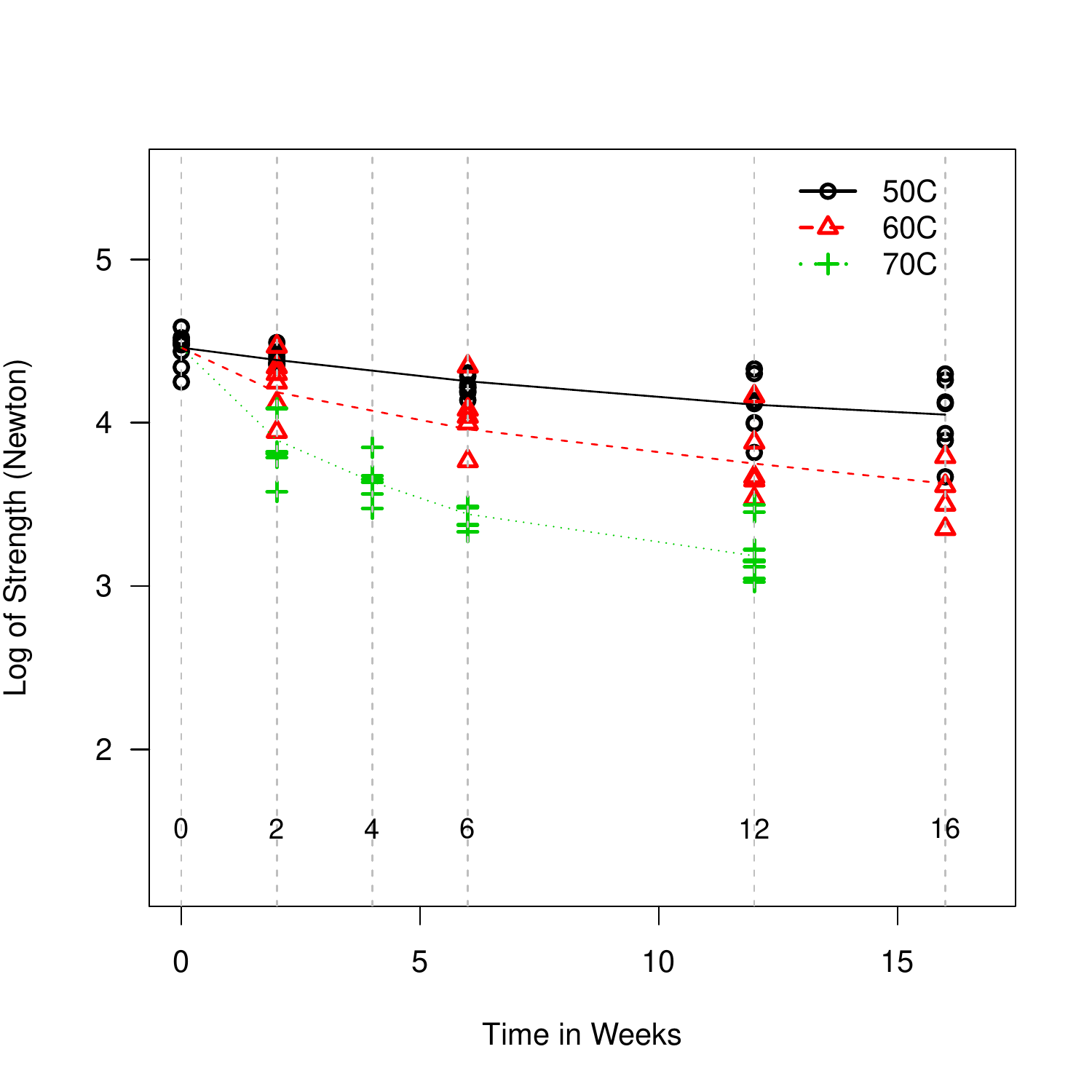}
        }
 \end{center}
 \caption{Fitted degradation paths of the Adhesive Bond B data.} \label{fig: AdhesiveBondB}
\end{figure}

\shortciteN{Escobaretal2003} discussed an experiment that measured the strength of an adhesive bond (Adhesive Bond B) over time. Eight units were measured at the beginning of the experiment under normal temperature to serve as the baseline strength. The remaining measurements were taken at selected weeks (2, 4, 6, 12, and 16) for three accelerated temperature levels ($50\degreeC$, $60\degreeC$, and $70\degreeC$). A scatter plot of Adhesive Bond B dataset is presented in Figure~\ref{fig: AdhesiveBondB}(a). The degradation model used by \shortciteN{Escobaretal2003} is
\begin{gather} \label{eqn: AdhesiveBondB parametric model}
y_{ijk}=\beta_{0}+\beta_{1}\exp(\beta_{2} x_{i} ) \tau_{j} + \varepsilon_{ijk},
\end{gather}
where $y_{ijk}$ is the strength of Adhesive Bond B in log Newtons, $\tau_{j}= \sqrt{\textrm{Week}_{j}}, x_{i}=-11605/$ $(\textrm{Temp}_{i}+273.15)$ is the Arrhenius-transformed temperature, and $\varepsilon_{ijk} \sim \mathrm{N}(0, \sigma^2)$. The estimates are $\betahat_{0}=4.4713$, $\betahat_{1}=-8.6384\times 10^{8}$, $\betahat_{2}=0.6364$ and $\sigmahat=0.1609$.

\subsubsection{Seal Strength Data} \label{sec: Seal Strength Data}
\begin{figure}
     \begin{center}
             \subfigure[Parametric model] {
        \includegraphics[width=0.47\textwidth]{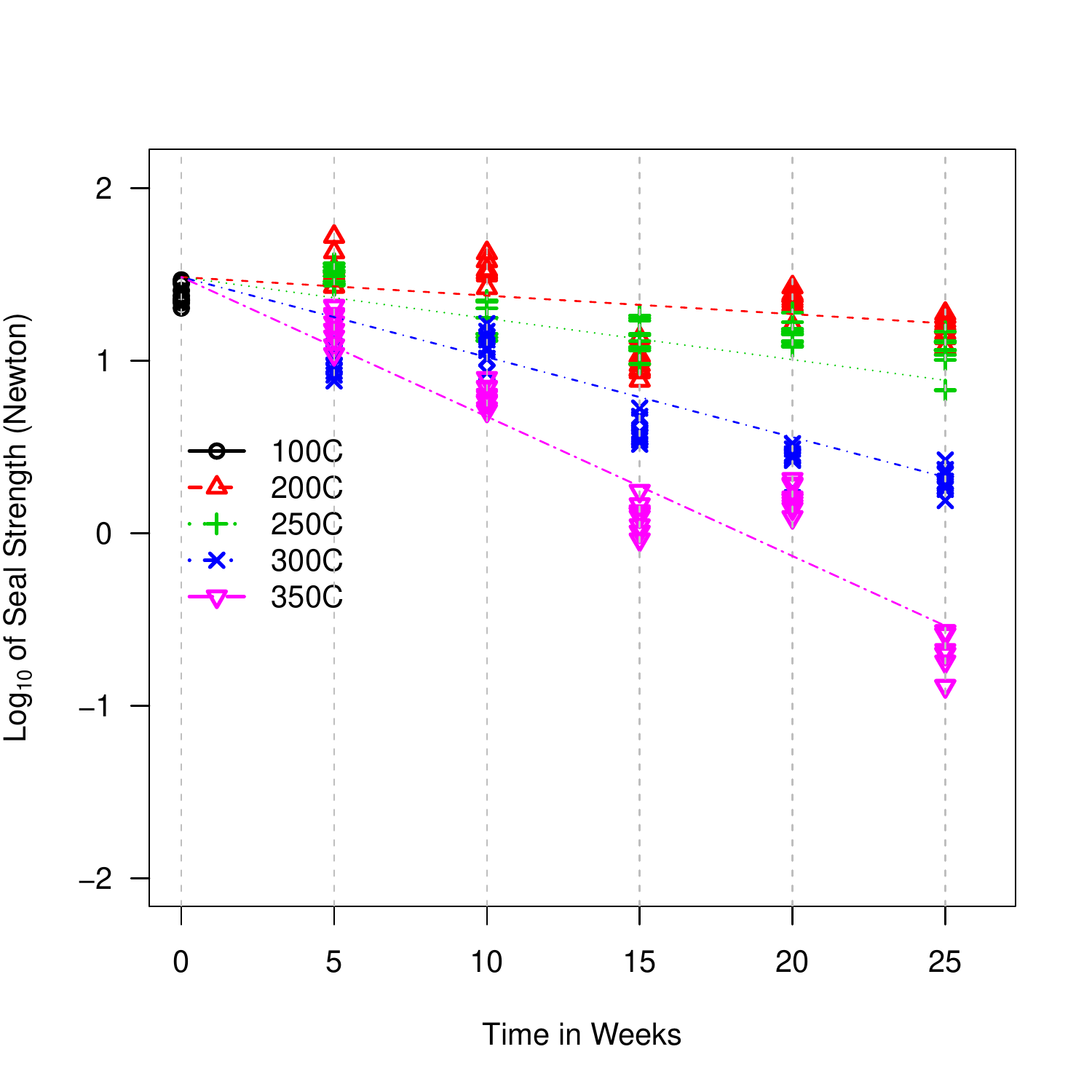}
                }
        \subfigure[Semi-parametric model] {
        \includegraphics[width=0.47\textwidth]{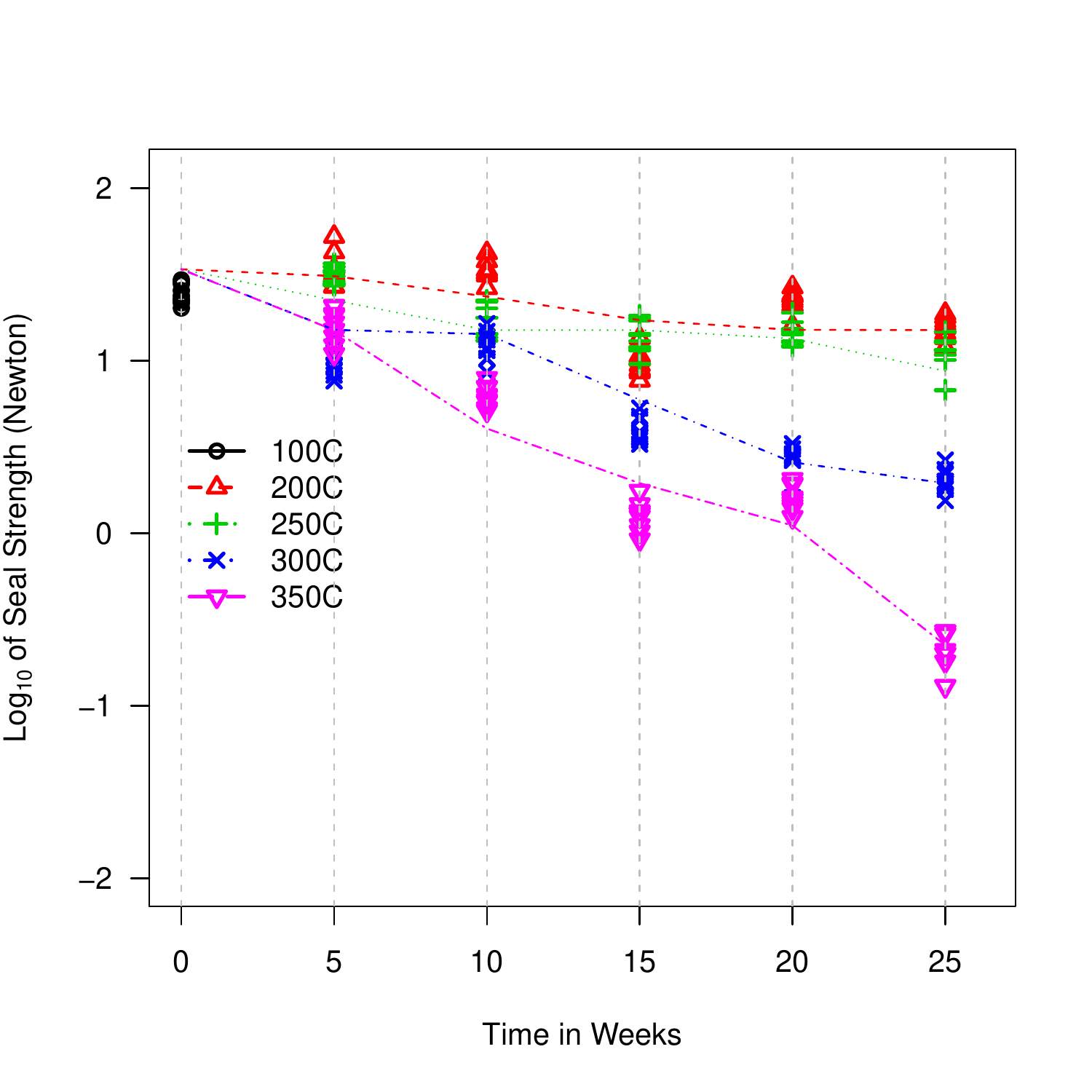}
        }
 \end{center}
 \caption{Fitted degradation paths of the Seal Strength data.} \label{fig: Sealstrength}
\end{figure}

Seal strength data were considered by \citeN{LiDoganaksoy2014}. At the start of the experiment, a batch of 10 seals were measured at the use temperature level of $100\degreeC$. A batch of 10 seal samples were then tested at selected weeks (5, 10, 15, 20, and 25) for four temperature levels ($200\degreeC$, $250\degreeC$, $300\degreeC$, and $350\degreeC$). A scatter plot of seal strength data is shown in Figure~\ref{fig: Sealstrength}(a). Though one would expect the seal strength to decrease under higher temperature, some batches of seal samples yielded higher strengths in later weeks compared with the initial measurements. This suggests a large batch-to-batch variability which must be incorporated into the model. Thus, \citeN{LiDoganaksoy2014} considered the following nonlinear mixed model:
\begin{gather} \label{eqn: Seal strength parametric model}
y_{ijk}=\beta_{0}-\beta_{1}\exp(\beta_{2} x_{i} ) \tau_{j} + \delta_{ij} + \varepsilon_{ijk},
\end{gather}
where $y_{ijk}$ is the $\log_{10}$ strength of seal sample, $\tau_{j}= \textrm{Week}_{j}$, and $x_{i}=-11605/(\textrm{Temp}_{i}+273.15)$. The random variable $\delta_{ij} \sim \mathrm{N}(0, \sigma_{\delta}^2)$ represents batch variability, $\varepsilon_{ijk} \sim \mathrm{N}(0, \sigma^2)$, and  $\delta_{ij}$ and $\varepsilon_{ijk}$ are independent. The estimates are $\betahat_{0}=1.4856$, $\betahat_{1}=47.2166$, $\betahat_{2}=0.3420$, $\sigmahat=0.1603$, and $\sigmahat_{\delta}=0.0793$.

\subsubsection{Adhesive Formulation~K Data} \label{sec: Adhesive Formulation K}
\begin{figure}
     \begin{center}
             \subfigure[Parametric model] {
        \includegraphics[width=0.47\textwidth]{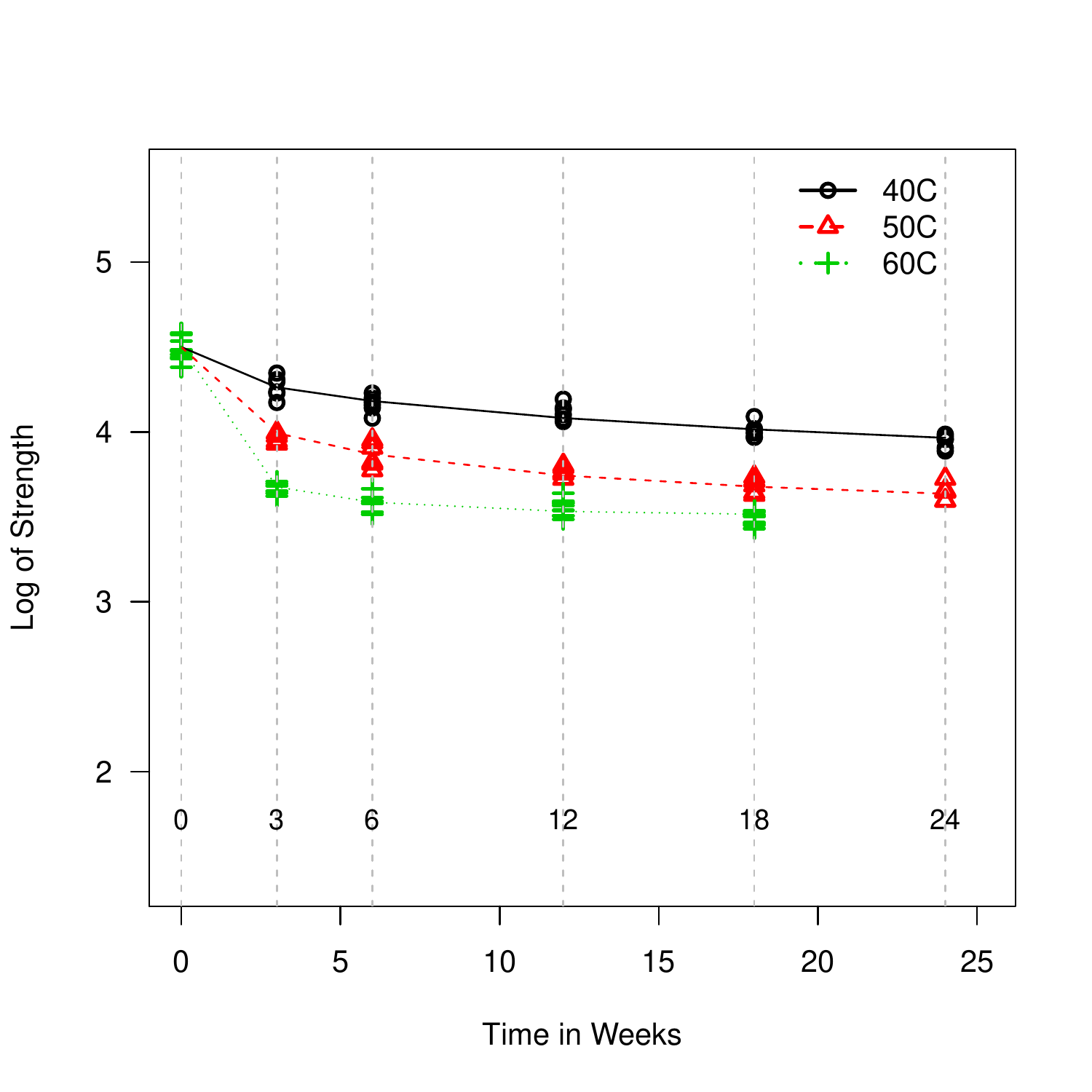}
                }
        \subfigure[Semi-parametric model] {
        \includegraphics[width=0.47\textwidth]{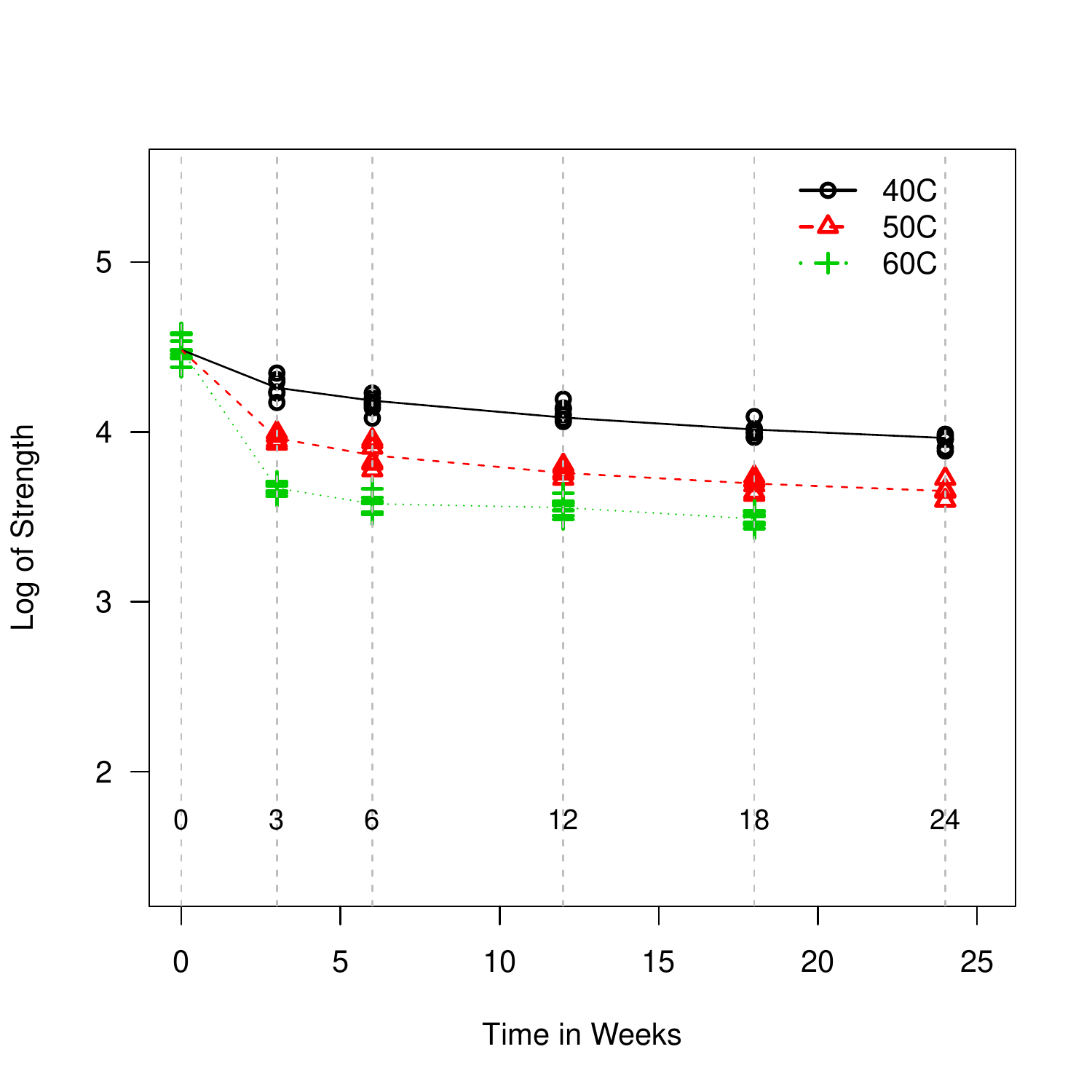}
        }
 \end{center}
 \caption{Fitted degradation paths of the Adhesive Formulation~K data.} \label{fig: AdheisiveFormulationK}
\end{figure}

A new adhesive (Formulation K) was developed and tested at 40$\degreeC$, 50$\degreeC$, and 60$\degreeC$. The strength of 10 units were measured at the beginning of the experiment and a specified number of samples were tested at 3, 6, 12, 18, and 24 weeks. Figure~\ref{fig: AdheisiveFormulationK}(a) is a scatter plot of the data. The nonlinear degradation model is
\begin{gather} \label{eqn: Adhesive Formulation K parametric model}
y_{ijk} = \log(90)+\beta_{0} (1-\exp\left\{-\beta_{1}  \exp\left[\beta_{2}(x_{i}-x_{2}) \right] \tau_{j} \right\}) + \varepsilon_{ijk},
\end{gather}
where $y_{ijk}$ is the strength of Adhesive Formulation K in log Newtons, $\tau_{j}= \sqrt{\textrm{Week}}_{j}$, $x_{i}=-11605/(\textrm{Temp}_{i}+273.15)$, $x_{2}=-11605/(50+273.15)$, and $\varepsilon_{ijk} \sim \mathrm{N}(0, \sigma^2)$. The estimates are $\betahat_{0}=-0.9978$, $\betahat_{1}=0.4091$, $\betahat_{2}=0.8371$, and $\sigmahat=0.0501$.

\subsection{Comparisons of Parametric and Semi-parametric Models} \label{sec: Parametric Fits and Semi-parametric Fits}
In order to assess the fit of the semi-parametric model, we applied it to each of the datasets and compared it with the corresponding parametric model chosen by the respective applications. We applied the knot selection technique in Section \ref{sec: Knots Selection} for each application. We also tested the significance of $\rho=0$ which informs the selection of the appropriate model. The parameter estimates and CI, as well as the MTTF at the normal use condition are presented in Tables \ref{tab: ests and CI of four applications} and \ref{tab: MTTF at normal use condition.}. The AIC defined in Section \ref{sec: Knots Selection} can also be used to compare the parametric and semi-parametric models. In the calculation of AIC,  the log-likelihood is the marginal log-likelihood for the parametric models.

\begin{table}
\centering
\caption{Parameter estimates and corresponding CI for the semi-parametric models for the three applications.}
\label{tab: ests and CI of four applications}
\begin{tabular}{c|c|ccc}\hline\hline
\multirow{2}{*}{Applications}  & \multirow{2}{*}{Parameter} & \multirow{2}{*}{Estimate} & \multicolumn{2}{c}{Quantile-based CI} \\\cline{4-5}
 & & &95\% lower & 95\% upper\\\hline
\multirow{2}{*}{Adhesive Bond B}        & $\beta$  & 1.3422 & 1.1071 & 1.6165 \\
                                        & $\sigma$ & 0.1537 & 0.1265 & 0.1787 \\ \hline
\multirow{3}{*}{Seal Strength}          & $\beta$  & 0.3235 & 0.2451 & 0.5194 \\
                                        & $\sigma$ & 0.1610 & 0.1192 & 0.1904 \\
                                        & $\rho$   & 0.7573 & 0.5465 & 0.8307 \\   \hline
\multirow{2}{*}{Adhesive Formulation K} & $\beta$  & 1.8221 & 1.6575 & 2.3658 \\
                                        & $\sigma$ & 0.0484 & 0.0419 & 0.0544 \\ \hline\hline
\end{tabular}
\end{table}

\begin{table}
\centering
\caption{Estimated MTTF at normal use condition based on parametric and semi-parametric models for the three applications (time in weeks).}
\label{tab: MTTF at normal use condition.}
\begin{tabular}{c|cccc}\hline\hline
\multirow{2}{*}{Applications}  & Failure   & Normal Use        & Parametric & Semi-parametric \\
                               & Threshold & Conditions  & Models      & Models         \\\hline
Adhesive Bond B        & 70\% &  30$\degreeC$ & 270 & 306 \\
Seal Strength          & 70\% & 100$\degreeC$ & 222 & 127 \\
Adhesive Formulation K & 70\% &  30$\degreeC$ &  68 &  92 \\\hline\hline
\end{tabular}
\end{table}

Table \ref{tab: loglikelihoods of parametric fits and semi-parametric fits} contains the log-likelihood values, \textit{edf}, and AIC for each model and dataset. For all three datasets, the  semi-parametric models possessed a lower AIC as compared to the parametric models. The fitted degradation paths for the parametric and semi-parametric models are presented in Figures \ref{fig: AdhesiveBondB}, \ref{fig: Sealstrength}, and \ref{fig: AdheisiveFormulationK}. All three figures show that the semi-parametric models provide a good fit to the data. We can see that the proposed model is flexible in fitting ADDT data from different applications.


\begin{table}
\centering
\caption{Log likelihood and  AIC values of parametric and semi-parametric models for the ADDT data from the three applications.} \label{tab: loglikelihoods of parametric fits and semi-parametric fits}
\begin{tabular}{c|ccc|ccc}\hline\hline
\multirow{2}{*}{Applications} & \multicolumn{3}{ c| }{Parametric Models}  & \multicolumn{3}{ c }{Semi-parametric Models} \\ \cline{2-7}
 & Loglik & \textit{df} & AIC  & Loglik & \textit{edf} & AIC \\\hline
Adhesive Bond B        & 34.9665  &  4 &           -61.9330 &  38.7264  & 5 & -67.4418 \\
Seal Strength          & 194.9907 &  5 &          -379.9814 & 199.7454  & 6 & -387.4909 \\
Adhesive Formulation K & 158.9508 &  4 &          -309.9016 & 163.9898  & 8 & -311.9797 \\ \hline\hline
\end{tabular}
\end{table}

\section{Conclusions and Areas for Future Work} \label{sec: Conclusion and Areas for Future Work}
In this paper, we describe a new semi-parametric degradation model for ADDT data based on monotone B-splines. We develop estimation and inference procedures for the proposed model as well as methods for selecting knot locations for the B-splines. Our simulation results indicate that the proposed estimation procedures for our semi-parametric model perform very well. Compared to parametric models, our semi-parametric approach is more flexible and can be applied to a wide range of applications and may be best suited as a generic method for ADDT data analysis for industrial standards. In addition, the semi-parametric model is more robust to model misspecification than a parametric model approach.

One key application of our semi-parametric model could be for test planning. A test plan based on this model would be general enough for application to a variety of materials and also allow for testing of different models. Our model can be served as a starting ground from which to test models against the data gathered rather than having to assume a given model prior to data collection. This would certainly serve as an interesting topic for future research.

The models considered here were solely scale-acceleration models. However, for certain types of products, a model with both scale and shape acceleration may describe the degradation path more appropriately. For example, \shortciteN{Tsaietal2013} considered a parametric model with both scale and the shape acceleration in test planning. Estimation and inference procedures for the semi-parametric model would certainly be more complex with the introduction of a shape acceleration parameter. It would be of great interest to pursue this in future research.


\bibliographystyle{chicago}
\bibliography{addt_ref}

\begin{thebibliography}{}

\bibitem[\protect\citeauthoryear{Bollaerts, Eilers, and Mechelen}{Bollaerts
  et~al.}{2006}]{BollaertsEilersMechelen2006}
Bollaerts, K., P.~H. Eilers, and I.~Mechelen (2006).
\newblock Simple and multiple p-splines regression with shape constraints.
\newblock {\em British Journal of Mathematical and Statistical
  Psychology\/}~{\em 59\/}(2), 451--469.

\bibitem[\protect\citeauthoryear{Carpenter, Goldstein, and Rasbash}{Carpenter
  et~al.}{2003}]{CarppenterGoldsteinrasbash2003}
Carpenter, J.~R., H.~Goldstein, and J.~Rasbash (2003).
\newblock A novel bootstrap procedure for assessing the relationship between
  class size and achievement.
\newblock {\em Applied Statistics\/}~{\em 52}, 431--443.

\bibitem[\protect\citeauthoryear{De~Boor}{De~Boor}{2001}]{Boor2001}
De~Boor, C. (2001).
\newblock {\em A practical guide to splines}.
\newblock New York: Springer-Verlag.

\bibitem[\protect\citeauthoryear{Efron and Tibshirani}{Efron and
  Tibshirani}{1993}]{EfronTibshirani1993}
Efron, B. and R.~Tibshirani (1993).
\newblock {\em An Introduction to the Bootstrap}.
\newblock FL: Boca Raton: Chapman and Hall/CRC.

\bibitem[\protect\citeauthoryear{Eilers and Marx}{Eilers and
  Marx}{1996}]{EilersMarx1996}
Eilers, P.~H. and B.~D. Marx (1996).
\newblock Flexible smoothing with b-splines and penalties.
\newblock {\em Statistical science\/}, 89--102.

\bibitem[\protect\citeauthoryear{Escobar, Meeker, Kugler, and Kramer}{Escobar
  et~al.}{2003}]{Escobaretal2003}
Escobar, L.~A., W.~Q. Meeker, D.~L. Kugler, and L.~L. Kramer (2003).
\newblock Accelerated destructive degradation tests: Data, models, and
  analysis.
\newblock {\em Mathematical and Statistical Methods in Reliability\/},
  319--338.

\bibitem[\protect\citeauthoryear{Fengler and Hin}{Fengler and
  Hin}{2014}]{Fengler2014}
Fengler, M.~R. and L.-Y. Hin (2014).
\newblock A simple and general approach to fitting the discount curve under
  no-arbitrage constraints.
\newblock {\em Available at SSRN:\/}.
\newblock http://ssrn.com/abstract=2478719.

\bibitem[\protect\citeauthoryear{Fritsch and Carlson}{Fritsch and
  Carlson}{1980}]{FritschCarlson1980}
Fritsch, F.~N. and R.~E. Carlson (1980).
\newblock Monotone piecewise cubic interpolation.
\newblock {\em SIAM Journal on Numerical Analysis\/}~{\em 17\/}(2), 238--246.

\bibitem[\protect\citeauthoryear{Gorjian, Ma, Mittinty, Yarlagadda, and
  Sun}{Gorjian et~al.}{2010}]{Gorjianetal2010}
Gorjian, N., L.~Ma, M.~Mittinty, P.~Yarlagadda, and Y.~Sun (2010).
\newblock A review on degradation models in reliability analysis.
\newblock In {\em Engineering Asset Lifecycle Management}, pp.\  369--384.
  Springer.

\bibitem[\protect\citeauthoryear{He and Shi}{He and Shi}{1998}]{HeShi1998}
He, X. and P.~Shi (1998).
\newblock Monotone b-spline smoothing.
\newblock {\em Journal of the American statistical Association\/}~{\em
  93\/}(442), 643--650.

\bibitem[\protect\citeauthoryear{Hofner, Kneib, and Hothorn}{Hofner
  et~al.}{2014}]{HofnerKneibHothorn2014}
Hofner, B., T.~Kneib, and T.~Hothorn (2014).
\newblock A unified framework of constrained regression.
\newblock {\em Statistics and Computing\/}, 1--14.

\bibitem[\protect\citeauthoryear{Hofner, M{\"u}ller, and Hothorn}{Hofner
  et~al.}{2011}]{Hofneretal2011}
Hofner, B., J.~M{\"u}ller, and T.~Hothorn (2011).
\newblock Monotonicity-constrained species distribution models.
\newblock {\em Ecology\/}~{\em 92\/}(10), 1895--1901.

\bibitem[\protect\citeauthoryear{Hong, Duan, Meeker, Stanley, and Gu}{Hong
  et~al.}{2015}]{Hongetal2015}
Hong, Y., Y.~Duan, W.~Q. Meeker, D.~L. Stanley, and X.~Gu (2015).
\newblock Statistical methods for degradation data with dynamic covariates
  information and an application to outdoor weathering data.
\newblock {\em Technometrics\/}~{\em 57}, 180--193.

\bibitem[\protect\citeauthoryear{Kanungo, Gay, and Haralick}{Kanungo
  et~al.}{1995}]{KanungoGayHaralick1995}
Kanungo, T., D.~M. Gay, and R.~M. Haralick (1995).
\newblock Constrained monotone regression of roc curves and histograms using
  splines and polynomials.
\newblock In {\em Image Processing, 1995. Proceedings., International
  Conference on}, Volume~2, pp.\  292--295. IEEE.

\bibitem[\protect\citeauthoryear{Leitenstorfer and Tutz}{Leitenstorfer and
  Tutz}{2007}]{LeitenstorferTutz2007}
Leitenstorfer, F. and G.~Tutz (2007).
\newblock Generalized monotonic regression based on b-splines with an
  application to air pollution data.
\newblock {\em Biostatistics\/}~{\em 8\/}(3), 654--673.

\bibitem[\protect\citeauthoryear{Li and Doganaksoy}{Li and
  Doganaksoy}{2014}]{LiDoganaksoy2014}
Li, M. and N.~Doganaksoy (2014).
\newblock Batch variability in accelerated-degradation testing.
\newblock {\em Journal of Quality Technology\/}~{\em 46}, 171--180.

\bibitem[\protect\citeauthoryear{Lu and Meeker}{Lu and
  Meeker}{1993}]{LuMeeker1993}
Lu, C.~J. and W.~Q. Meeker (1993).
\newblock Using degradation measures to estimate a time-to-failure
  distribution.
\newblock {\em Technometrics\/}~{\em 34}, 161--174.

\bibitem[\protect\citeauthoryear{Meeker, Escobar, and Lu}{Meeker
  et~al.}{1998}]{MeekerEscobarLu1998}
Meeker, W.~Q., L.~A. Escobar, and C.~J. Lu (1998).
\newblock Accelerated degradation tests: modeling and analysis.
\newblock {\em Technometrics\/}~{\em 40\/}(2), 89--99.

\bibitem[\protect\citeauthoryear{Meeker, Hong, and Escobar}{Meeker
  et~al.}{2011}]{MeekerHongEscobar2011}
Meeker, W.~Q., Y.~Hong, and L.~A. Escobar (2011).
\newblock Degradation models and data analyses.
\newblock In {\em Encyclopedia of Statistical Sciences}. Wiley.

\bibitem[\protect\citeauthoryear{Meyer}{Meyer}{2012}]{Meyer2012}
Meyer, M. (2012).
\newblock Constrained penalized splines.
\newblock {\em Canadian Journal of Statistics\/}~{\em 40}, 190--206.

\bibitem[\protect\citeauthoryear{Meyer}{Meyer}{2008}]{Meyer2008}
Meyer, M.~C. (2008).
\newblock Inference using shape-restricted regression splines.
\newblock {\em The Annals of Applied Statistics\/}~{\em 2}, 1013--1033.

\bibitem[\protect\citeauthoryear{Nelson}{Nelson}{1990}]{Nelson1990}
Nelson, W.~B. (1990).
\newblock {\em Accelerated testing: statistical models, test plans, and data
  analysis}.
\newblock John Wiley \& Sons.

\bibitem[\protect\citeauthoryear{Ramsay}{Ramsay}{1988}]{Ramsay1988}
Ramsay, J.~O. (1988).
\newblock Monotone regression splines in action.
\newblock {\em Statistical science\/}~{\em 3}, 425--441.

\bibitem[\protect\citeauthoryear{Tsai, Tseng, Balakrishnan, and Lin}{Tsai
  et~al.}{2013}]{Tsaietal2013}
Tsai, C.-C., S.-T. Tseng, N.~Balakrishnan, and C.-T. Lin (2013).
\newblock Optimal design for accelerated destructive degradation tests.
\newblock {\em Quality Technology and Quantitative Management\/}~{\em 10},
  263--276.

\bibitem[\protect\citeauthoryear{UL746B}{UL746B}{2013}]{UL746B}
UL746B (2013).
\newblock {\em Polymeric Materials - Long Term Property Evaluations, UL 746B}.
\newblock Underwriters Laboratories, Incorporated.

\bibitem[\protect\citeauthoryear{Vaca-Trigo and Meeker}{Vaca-Trigo and
  Meeker}{2009}]{VacaTrigoMeeker2009}
Vaca-Trigo, I. and W.~Q. Meeker (2009).
\newblock A statistical model for linking field and laboratory exposure results
  for a model coating.
\newblock In J.~Martin, R.~A. Ryntz, J.~Chin, and R.~A. Dickie (Eds.), {\em
  Service Life Prediction of Polymeric Materials}, Chapter~2. NY: New York:
  Springer.

\bibitem[\protect\citeauthoryear{Wang, Meyer, and Opsomer}{Wang
  et~al.}{2013}]{WangMeyerOpsomer2013}
Wang, H., M.~C. Meyer, and J.~D. Opsomer (2013).
\newblock Constrained spline regression in the presence of {AR(p)} errors.
\newblock {\em Journal of Nonparametric Statistics\/}~{\em 25}, 809--827.

\bibitem[\protect\citeauthoryear{Xu, Hong, and Jin}{Xu
  et~al.}{2015}]{XuHongJin2015}
Xu, Z., Y.~Hong, and R.~Jin (2015).
\newblock Nonlinear general path models for degradation data with dynamic
  covariates.
\newblock {\em Applied Stochastic Models in Business and Industry, in press,
  doi: 10.1002/asmb.2129\/}.

\bibitem[\protect\citeauthoryear{Ye and Xie}{Ye and Xie}{2015}]{YeXie2015}
Ye, Z. and M.~Xie (2015).
\newblock Stochastic modelling and analysis of degradation for highly reliable
  products.
\newblock {\em Applied Stochastic Models in Business and Industry\/}~{\em 31},
  16--32.

\bibitem[\protect\citeauthoryear{Ye, Xie, Tang, and Chen}{Ye
  et~al.}{2014}]{Yeetal2014}
Ye, Z.-S., M.~Xie, L.-C. Tang, and N.~Chen (2014).
\newblock Semiparametric estimation of {Gamma} processes for deteriorating
  products.
\newblock {\em Technometrics\/}~{\em 56}, 504--513.

\end{thebibliography}
\end{document}